\documentclass[aps,prx,twocolumn,showpacs,longbibliography]{revtex4-1}
\usepackage{amsmath}
\usepackage{graphicx}
\usepackage[colorlinks,linkcolor=blue,citecolor=blue,urlcolor=green]{hyperref}

\newcommand{\iu}{\mathrm{i}}
\newcommand{\dd}{\mathrm{d}}

\renewcommand{\vec}[1]{\mathbf{#1}}
\newcommand{\gvec}[1]{\boldsymbol{#1}}
\newcommand{\en}{\varepsilon}
\newcommand{\hc}{\hat{c}}
\newcommand{\hcd}{\hat{c}^\dagger}

\begin{document}

\title{Tracing the nonequilibrium topological state of Chern insulators}

\author{Michael Sch\"uler}
\author{Philipp Werner}
\affiliation{Department of Physics, University of Fribourg, 1700 Fribourg, Switzerland}

\begin{abstract}
  Chern insulators exhibit fascinating properties which originate from
  the topologically nontrivial state characterized by the Chern
  number.  How these properties change if the system is quenched
  between topologically distinct phases has however not been
  systematically explored.  In this work, we investigate the quench
  dynamics of the prototypical massive Dirac model for topological
  insulators in two dimensions. We consider both dissipation-less
  dynamics and the effect of electron-phonon interactions, and ask how
  the transient dynamics and nonequilibrium steady states affect
  simple observables.  Specifically, we discuss a time-dependent
  generalization of the Hall effect and the dichroism of the
  photoexcitation probability between left and right circularly
  polarized light. We present optimized schemes based on these
  observables, which can reveal the evolution of the topological state
  of the quenched system.
\end{abstract}


\maketitle

\section{Introduction}

Topologically nontrivial phases of matter are a subject of intense
current research~\cite{hasan_colloquium:_2010,moore_birth_2010}. They
exhibit a range of intriguing and potentially useful properties, such
as the quantum anomalous Hall (QAH) effect.  The usual notion is that
the intrinsic topology of a system cannot be altered by local
perturbations, which results in the protection of certain properties
due to time-reversal symmetry. In particular, this effect leads to
stable surface or edge states and their extraordinary transport
properties.

The relation between the topological phase and corresponding
observables is still under active investigation. Originally, such a
correspondance has been established in non-interacting systems in
equilibrium via the Thouless-Kohmoto-Nightingale-Tijs
formula~\cite{thouless_quantized_1982}. In this case, the Chern number
$\mathcal{C}_n$ of each band (labelled by $n$) fully characterizes the
QAH effect: the Hall conductivity amounts to
$\sigma_{xy} = (e^2/h) \sum_{n\in \mathrm{occ}} \mathcal{C}_n$. The
bulk-insulating system (characterized by $\sigma_{xx} = 0$ and
$\sigma_{xy}\ne 0$) is then called a Chern or quantum Hall insulator
(QHI). The Chern number, on the other hand, is obtained via the Berry
curvature from the wave-function $|\phi_{\vec{k}n}\rangle$ directly --
a quantity that is, strictly speaking, available for non-interacting
systems (or within mean-field treatments) only.  A possible extension
in the context of many-body perturbation theory can be obtained by
constructing an effective Hamiltonian involving the self-energy at
zero frequency~\cite{wang_simplified_2012}, which allows to study the
interplay of topological properties and correlation effects in 
QHIs~\cite{budich_fluctuation-driven_2013,amaricci_first-order_2015,
  kumar_interaction-induced_2016}. This approach is based on the
connection of the Chern number to the winding
number~\cite{wang_topological_2010}.  Alternatively, topological
states can be classified by studying the response of a system to
external gauge fields~\cite{frohlich_gauge_2013}.  While these
approaches work for noninteracting as well as for interacting
electrons, the underlying assumption is that the system is in its
\emph{ground state}. Hence the established concepts are not
necessarily applicable to finite temperature or \emph{nonequilibrium}
scenarios, which involve excited states.

This is particularly true for global perturbations such as quenches of
the Hamiltonian parameters. For instance, a straightforward definition
of a time-dependent Chern number $\mathcal{C}_n(t)$ from the
time-evolving wave-functions $|\phi_{\vec{k}n}(t)\rangle$ of a 
non-interacting system will remain constant under unitary evolution.
On the other hand, the Hall conductivity $\sigma_{xy}$ can change,
e. g. after a quench, which means that it is generally {\it not} identical
to the Chern number (up to the prefactor $e^2/h$), in contrast to the
equilibrium case~\cite{caio_quantum_2015,wang_universal_2016,
  unal_nonequilibrium_2016}. Similarly, the
bulk-boundary-correspondance might be lost after a
quench~\cite{schmitt_universal_2017}. It is thus a relevant task to
identify experimentally accessible quantities which allow to trace the
nonequilibrium evolution of topologically nontrivial systems.

In this work, we investigate different schemes that enable us to
study the nonequilibrium and \emph{transient} dynamics of QHIs. We
focus on (i) the time-resolved Hall effect using appropriately shaped
electromagnetic pulses, and (ii) the photoabsorption asymmetry with
respect to left/right circularly polarized light -- a novel approach
which has been suggested recently~\cite{tran_probing_2017}. Both
methods are based on directly observable quantities and are thus
well suited for the study of (effectively) noninteracting as well as correlated
and/or dissipative systems. We demonstrate their applicability by considering the
well-known massive Dirac model (MDM) on a square
lattice \cite{qi_topological_2008}, which captures~\cite{liu_quantum_2008} the topoligical
phase transition in HgTe quantum
wells~\cite{bernevig_quantum_2006} as a generic example.
We focus on quench dynamics and demonstrate
how a transition between phases of distinct topological character
manifests itself in these observables. Furthermore, we study the influence of 
dissipation due to electron-phonon (el-ph) coupling on the transient dynamics
to demonstrate the robustness of the proposed schemes.

\section{Model}

As a paradigm model for two-dimensional systems we consider the
massive Dirac model on a square lattice. The electronic Hamiltonian reads
\begin{align}
  \label{eq:Hel}
  \hat{H}_\mathrm{el}  = \sum_{\vec{k}\in\mathrm{BZ}} \sum_{ab}
  [h_\text{el}(\vec{k})]_{ab} \hcd_{\vec{k} a} \hc_{\vec{k} b} \ ,
\end{align} 
where the $\vec{k}$-dependent single-particle Hamiltonian $[h_\mathrm{el}(\vec{k})]_{ab} =
\langle \vec{k} a | \hat{h}_\mathrm{el}(\vec{k}) | \vec{k} b \rangle$ has the generic
form
$\hat{h}_\mathrm{el}(\vec{k}) = \sum_{\alpha=x,y,z} d_{\alpha}(\vec{k})
\hat{\sigma}^\alpha$. Here, the $\hat{\sigma}^\alpha$ denote the
pseudo-spin operators with respect to the underlying bands. The coefficients
$d_{\alpha}(\vec{k})$ are defined by 
\begin{align}
  \label{eq:dvec}
  d_x(\vec{k}) &= \lambda \sin(k_x a) \ , \ d_y(\vec{k}) = \lambda
  \sin(k_y a) \ , \nonumber\\  d_z(\vec{k}) &= -2T_0 \left( \cos(k_x a) +
                                     \cos(k_y a) \right) -M \ .
\end{align} 
The eigenstates of the single-particle Hamiltonian are denoted by
$\hat{h}_\mathrm{el}(\vec{k}) |\phi_{\vec{k}n} \rangle = \en_n(\vec{k})
|\phi_{\vec{k}n} \rangle $.  

Note that we
limit ourselves to a spin-restricted model here, as the
Hamiltonian~\eqref{eq:Hel} is spin-independent.
As shown in Ref.~\onlinecite{liu_quantum_2008}, in HgTe quantum wells
-- which are
well modelled by Eqs.~\eqref{eq:Hel} and \eqref{eq:dvec} -- doping
with manganese allows to shift the spin up/down bands in such a way that
only one spin channel remains important. Here we assume such a
situation and thus focus on the charge QAH effect instead of the usual
quantum spin Hall effect (QSH). 

It is straightforward to see that diagonalizing the
Hamiltonian~\eqref{eq:Hel} gives rise to a trivial band insulator (BI)
for $M<-4|T_0|$, while the system corresponds to a topological
insulator (TI)
\footnote{Even though the term topological insulator (TI) refers to a more
  general concept than the QAH insulator, we use the abbreviation TI
  throughout the text.}
 for $0>M>-4|T_0|$.
As usual, the topological character
of the bands can be determined by the Chern number $\mathcal{C}_n$,
which is defined (for each band $n=1,2$, respectively) by the integral over
the Berry curvature
\begin{align}
  \label{eq:chern}
  \mathcal{C}_n = \frac{1}{2\pi} \int\! \dd \vec{k}\,
  \left(\frac{\partial A^{(n)}_y}{\partial k_x} - \frac{\partial
  A^{(n)}_x}{\partial k_y} \right) \ .
\end{align}
Here, $\vec{A}^{(n)}(\vec{k})=-\iu \langle
\phi_{\vec{k}n}|\nabla_{\vec{k}} \phi_{\vec{k}n}\rangle$
is the Berry connection 
corresponding to the upper ($n=1$)
or lower ($n=2$) band.

\subsection{Quench dynamics\label{subsec:quenchdyn}}

A quench from the BI into the TI phase (or vice-versa) offers insight
into the underlying topological properties. For instance, the insulating
state with nontrivial Chern number cannot be altered under unitary
time evolution (which preserves time-reversal symmetry) -- hence, the
question arises to which state the system is driven. Furthermore,
quenches may induce dynamical phase transitions, which have recently
been studied for the MDM~\cite{heyl_dynamical_2017}. Quenches can be
realized, for instance, by photodoping pulses in interacting
electronic systems, leading to transient band
shifts~\cite{eckstein_photoinduced_2013,golez_nonequilibrium_2017-1},
or by changing the strength of the periodic driving in Floquet topological
insulators~\cite{lindner_floquet_2011,dalessio_dynamical_2015}.

Here we study quenches of the gap parameter $M$ between the
values $M_\mathrm{TI}=-3|T_0|$ and $M_\mathrm{BI}=-5|T_0|$. The band
hybridization is fixed at $\lambda=0.2 |T_0|$ in what follows.
The transition from phase A ($M = M_\mathrm{TI/BI}$) to B
($M = M_\mathrm{BI/TI}$) is triggered by a softened ramp of the form
\begin{align}
  \label{eq:Hquench}
  \hat{H}_\mathrm{el}(t) = (1-\alpha(t))\hat{H}^{\mathrm{A}}_\mathrm{el} + \alpha(t)\hat{H}^{\mathrm{B}}_\mathrm{el} ,
\end{align}
with $\alpha(t)=1-\cos(\pi(t-t_\mathrm{q})/T_\mathrm{q})$, defined by
the quench time $t_\mathrm{q}$ and the duration $T_\mathrm{q}$
($t_\mathrm{q} < t < t_\mathrm{q} + T_\mathrm{q}$).

In the absence of any further interactions, the
dynamics can be captured by solving the time-dependent Schr\"odinger
equation
\begin{align}
  \label{eq:tdse1}
  \iu \frac{\dd}{\dd t} |\psi_{\vec{k}n}(t) \rangle =
  \hat{h}_\mathrm{el}(t)  |\psi_{\vec{k}n}(t) \rangle  \ , \
  |\psi_{\vec{k}n}(t=0) \rangle = \delta_{n,1} |\phi_{\vec{k}1}\rangle
  \ ,
\end{align}
where we have assumed half filling. The time-dependent
single-particle Hamiltonian $\hat{h}_\mathrm{el}(t) $ is defined in
analogy to Eq.~\eqref{eq:Hquench}. As mentioned above, a
straightforward generalization of the definition of the Chern number~\eqref{eq:chern} as
\begin{align}
  \mathcal{C}_n(t) = \frac{1}{2\pi} \int\! \dd \vec{k}\,
  \left(\frac{\partial A^{(n)}_y(t)}{\partial k_x} - \frac{\partial
  A^{(n)}_x(t)}{\partial k_y} \right) 
\end{align}
with the time-dependent Berry connection
$\vec{A}^{(n)}(\vec{k},t)=-\iu \langle
\psi_{\vec{k}n}(t)|\nabla_{\vec{k}}\psi_{\vec{k}n}(t)\rangle$ will be
invariant under any unitary time
evolution~\cite{caio_quantum_2015,schmitt_universal_2017}. As discussed by Wang \emph{et
al.} (ref.~\onlinecite{wang_universal_2016}), the Hall conductance
may however undergo a change. Assuming that decoherence effects have
suppressed the off-diagonal elements of the density matrix after
sufficiently long time, the steady-state Hall conductance can be
defined as
\begin{align}
  \label{eq:hallneq}
  \sigma_{xy} = \frac{e^2}{2\pi h} \sum_n  \int\! \dd \vec{k}\, f_n(\vec{k})
  \left(\frac{\partial A^{(n),\mathrm{B}}_y}{\partial k_x} - \frac{\partial
  A^{(n),\mathrm{B}}_x}{\partial k_y} \right) \ .
\end{align}
Here, $\vec{A}^{(n), \mathrm{B}}(\vec{k})$ stands for the Berry
connection of the \emph{post-quench} Hamiltonian
$\hat{h}^\mathrm{B}_\mathrm{el}$, whereas $f_n(\vec{k})$ is the
occupation with respect to the post-quench band structure.  For the
two-band MDM, $f_n(\vec{k})$ is easily expressed in terms of the
overlap of the pre- and post-quench basis and does not depend on the
quench details.

\subsection{Electron-phonon coupling\label{subsec:elph}}

\begin{figure}[t]
  \includegraphics[width=\columnwidth]{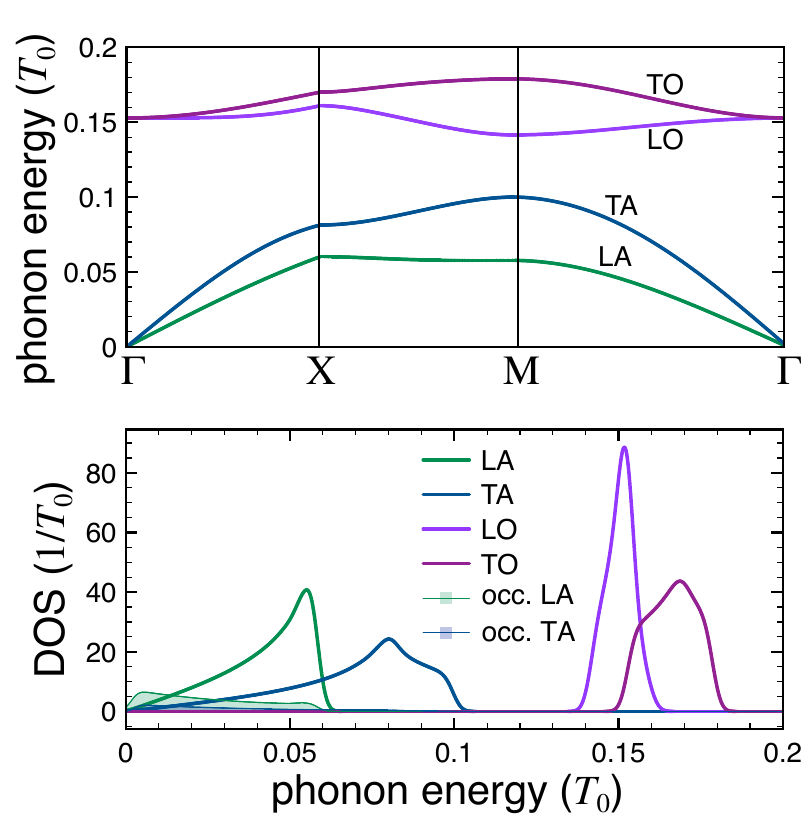}
  \caption{Upper panel: dispersion of the longitudinal acoustic (LA),
    transverse acoustic (TA), longitudinal optical (LO) and
    transverse optical (TO) phonon modes. The energy is measured in
    units of the hopping constant $T_0$. Lower panel: corresponding
    density of states (DOS) along with the occupation-weighted DOS for
    the LA and TA modes (filled curves). \label{fig:phonons}}
\end{figure}

The nonequilibrium Hall conductance~\eqref{eq:hallneq} assumes the
system to have lost the coherences due to environmental coupling;
otherwise, the coherent oscillations after excitations induced by the
quench hamper a unambiguous definition of $\sigma_{xy}$. The most important
intrinsic source for such dephasing effects in real systems is
(besides structural defects) el-ph coupling. Since the
coupling to the lattice vibrations entails -- besides the decoherence
effects -- dissipative population dynamics, the steady state 
reached by the system is -- in general -- different from the above quench
scenario~\cite{wolff_dissipative_2016}. Therefore, we treat the el-ph coupling
explicitly by extending the Hamiltonian to 
\begin{align}
  \label{eq:Htot}
  \hat{H}(t) = \hat{H}_\mathrm{el}(t) + \hat{H}_{\mathrm{el-ph}} +
  \hat{H}_{\mathrm{ph}} \ .
\end{align}
For the interaction term, we use the Fr\"ohlich
coupling~\cite{van_leeuwen_first-principles_2004} in two dimensions:
\begin{align}
  \label{eq:Helph}
  \hat{H}_{\mathrm{el-ph}} = \frac{\gamma}{\sqrt{N_k}} \sum_{\mu \vec{q}}
  \vec{u}_{\mu\vec{q}} \cdot \frac{\vec{q}}{q} \sum_{\vec k}
  \hcd_{\vec{k}n} \hc_{\vec{k}-\vec{q} n} \hat{Q}_{\mu \vec{q}} \ .
\end{align}
Here, $\gamma$ is a constant determining the overall coupling strength
and $N_k$ denotes the number of $\vec{k}$ points. The phonon modes
$\mu$ with momentum $\vec{q}$ are represented by the corresponding
mode vector $\vec{u}_{\mu \vec{q}}$ and the coordinate (momentum)
operators $ \hat{Q}_{\mu \vec{q}}$ ($\hat{P}_{\mu \vec{q}}$). The
latter define the phonon Hamiltonian by
\begin{align}
  \label{eq:Hph}
  \hat{H}_{\mathrm{ph}} = \sum_{\mu \vec{q}} \frac{\omega_{\mu}(\vec{q})}{2}
  \left(\hat{P}^\dagger_{\mu \vec{q}}\hat{P}_{\mu \vec{q}} +
  \hat{Q}^\dagger_{\mu \vec{q}}\hat{Q}_{\mu \vec{q}} \right) \ .
\end{align}
To compute the phonon modes in the square lattice, we assume a
diatomic basis characterized by masses $M_{1,2}$ and three force
constants $c_{1,2,3}$ describing the nearest neighbour, next-nearest
neighbour and diagonal interaction, respectively. Diagonalizing the
dynamical matrix yields two acoustic and two optical (longitudinal
and transverse) phonon modes. The parameters are chosen to produce a
similar phonon dispersion $\omega_\mu(\vec{q})$ as is known for
HgTe~\cite{ouyang_first-principles_2015} -- up to an overall scaling
constant. We have slightly increased the phonon energies to make the
effects due to the el-ph coupling more visible. However, it
is important to note that we stay in the realistic parameter regime
where only intraband transitions can be induced by scattering from
phonons. The phonon dispersion and the corresponding density of states
(DOS) is shown in Fig.~\ref{fig:phonons}. In what follows, we assume
low temperatures by fixing the inverse temperature at
$\beta = 40 |T_0|^{-1}$, such that only the acoustic phonons around
the $\Gamma$ point are thermally activated.

It should be mentioned that the role of the el-ph coupling in
topological insulators, in particular at the surface, is a topic of
recent discussions. While some works point out the strong influence of
el-ph coupling for inelastic
scattering~\cite{giraud_electron-phonon_2011,giraud_electron-phonon_2012,
costache_fingerprints_2014}
and for intraband relaxation of photoexcited
systems \cite{hatch_stability_2011}, other measurements suggest weak
el-ph interaction effects~\cite{pan_measurement_2012}. Here, we take a
different angle and treat the coupling strength $\gamma$ as a
parameter.

\subsection{Equations of motion in the presence of Electron-phonon
  interactions\label{subsec:eomelph}}

The full Hamiltonian~\eqref{eq:Htot} constitutes an interacting 
electron-boson model. The numerically exact solution is out of reach
for the typical number of points in reciprocal space. Furthermore,
since the phonon energies $\omega_{\mu}(\vec{q})$ are much smaller
than the electronic energy scale, a weak-coupling treatment is
suitable. In this context, the nonequilibrium Green's function (NEGF)
approach in its time-dependent formulation has become an important
tool recently \cite{sentef_examining_2013,kemper_mapping_2013,
  murakami_interaction_2015,sakkinen_many-body_2015-2,
  murakami_damping_2016,schuler_time-dependent_2016,
  tuovinen_phononic_2016}. However, most of the approaches resort to a
local approximation to the el-ph interaction. This
approximation excludes intraband transitions, which are, as discussed
above, the only available relaxation channel in our case. Therefore,
a momentum-dependent treatment of the phonons is required, which
increases the computational costs of the NEGF method
considerably. Since neither renormalization effects of the electronic
structure due to the coupling to the phonons nor the backaction on the
phononic degrees of freedom is of particular significance for the
relaxation in the MDM, one can employ a simplified theory based on a
master-equation approach within the Markovian
approximation~\cite{breuer_theory_2002}. The usual Lindblad equation,
however, is formulated in terms of the {\it many-body} density
matrix~\cite{usenko_femtosecond_2016-1}. For calculating the time
evolution of the {\it single-particle} density matrix,
$\rho_{ij}(\vec{k};t)= \langle \hcd_{\vec{k} i}(t) \hc_{\vec{k}j}(t)
\rangle $, additional conditions are required, reflecting not only the
overall particle conservation, but the fermionic statistics, as
well. This can be achieved by extending the linear Lindblad equation to
a nonlinear master
equation~\cite{rosati_derivation_2014,rosati_electron-phonon_2015}. Alternatively,
equivalent master equations can also be obtained from with NEGF
formulation within the generalized
Kadanoff-Baym ansatz~\cite{schlunzen_nonequilibrium_2016} and applying
the Markovian approximation~\cite{langreth_derivation_1991}.

Within the master-equation approach, the equation of motion (EOM) for
the single-particle density matrix reads
\begin{align}
  \label{eq:eomrho1}
  \frac{\dd}{\dd t} \gvec{\rho}(\vec{k};t) = -\iu \left[
  \vec{h}_\mathrm{el}(\vec{k};t), \gvec{\rho}(\vec{k};t)  \right] 
  + \vec{I}(\vec{k} ; t) \ ,
\end{align}
where we have employed a more compact matrix notation. Besides the
unitary time evolution captured by the first term, the el-ph
coupling is described by the scattering
term~\cite{rosati_derivation_2014}
\begin{align}
  \label{eq:scattterm}
  I_{ij}(\vec{k}; t) &= \frac12 \sum_{\mu \vec{q}} \sum_{k l m} \Big[ 
          \bar{\rho}_{ik}(\vec{k};t)\mathcal{R}^{\mu
                       \vec{q}}_{kl,jm}(\vec{k},\vec{k}-\vec{q})\rho_{l
                       m}(\vec{k}-\vec{q};t ) \nonumber\\
&\quad - \bar{\rho}_{kl}(\vec{k}+\vec{q};t) \mathcal{R}^{\mu
                       \vec{q}}_{ki,lm}(\vec{k}+\vec{q},\vec{k}) \rho_{mj}(\vec{k},t)
\Big] \ .
\end{align}
Here,
$\bar{\rho}_{ij}(\vec{k};t) = \delta_{ij} - \rho_{ij}(\vec{k};t)$ is the 
density matrix of the hole states, while
$\mathcal{R}^{\mu \vec{q}}_{kl,jm}(\vec{k}_1,\vec{k}_2)$ denotes
the scattering rate which determines the transition probability. As
the scattering events are limited to intraband transitions
governed by momentum (and energy) conservation, the scattering rates
simplify to~\cite{rosati_electron-phonon_2015}
\begin{align}
  \label{eq:scattrates}
  \mathcal{R}^{\mu \vec{q}}_{kl,jm}(\vec{k}_1,\vec{k}_2) = \delta_{\vec{k}_1,\vec{k}_2+\vec{q}}
  \delta_{k j} \delta_{lm}  R^{\mu \vec{q}}_{j l}(\vec{k}_1,\vec{k}_2)
\end{align}
with 
\begin{align}
  \label{eq:scattratessimple}
  R^{\mu \vec{q}}_{j l}(\vec{k}_1,\vec{k}_2) = \frac{\gamma^2}{N_k}\frac{1}{q^2}\left|\vec{u}_{\mu
  \vec{q}}\cdot \vec{q} \right|^2 D^>_{\mu \vec{q}}(\en_l(\vec{k}_2)
  -\en_j(\vec{k}_1) ) \ .
\end{align}
Here, the spectral function of the unoccupied phonon modes (greater
Green's function in the NEGF context) is defined by
\begin{align}
  \label{eq:phonongreater}
  D^>_{\mu \vec{q}} (\omega) = 2\pi \left( N_B(\omega) +1  \right)
  \left[\delta(\omega- \omega_\mu(\vec{q})) - \delta(\omega+
  \omega_\mu(\vec{q}))  \right] \ ,
\end{align}
where $N_B(\omega)$ denotes the Bose distribution. In practice, the
Dirac-$\delta$ functions in Eq.~\eqref{eq:phonongreater} are replaced
by Gaussians with a small broadening parameter $\eta$.

Note that the definition of the scattering
terms~\eqref{eq:scattterm}--\eqref{eq:scattratessimple} depends on the
electronic eigenenergies explicitly. The EOM~\eqref{eq:eomrho1} thus
has to be solved in the eigenbasis of
$\vec{h}_{\mathrm{el}}(\vec{k};t=0)$. For the quench scenario, this
means the electronic energies in Eq.~\eqref{eq:scattratessimple}
correspond to the pre-quench Hamiltonian.

The algorithm which we employed for the numerical solution of the master
equation~\eqref{eq:eomrho1} is presented in
appendix~\ref{app:mastereq}.

\section{Observables in equilibrium\label{sec:obseq}}

Let us now proceed by defining the observables which will be used to trace the
nonequilibrium dynamics. We first consider the equilibrium case, where the
system is either in the band insulating or QHI phase and then
discuss how to extend the schemes to the time-dependent case.

\subsection{Circular asymmetry\label{subsec:circasym}}

As suggested by Tran \emph{et al.}
(Ref.~\onlinecite{tran_probing_2017}), the depletion rate
$\Gamma^{(+)}_n(\omega)$ ($\Gamma^{(-)}_n(\omega)$) upon irradiation
of left (right) circularly polarized light with frequency $\omega$
yields a direct measure of the Chern number of band $n$. In
particular, the frequency-integrated asymmetry signal
$\int\!\dd \omega \, (\Gamma^{(+)}_n(\omega)-\Gamma^{(-)}_n(\omega))$
is proportional to $\mathcal{C}_n$. This property was derived for a
noninteracting system; however, it is generic and can also
be exploited in interacting systems. A closely related effect is the
pronounced polarization dependence observed in angle-resolved
photoemission (ARPES) from graphene, mapping out the Berry phase of
the individual bands \cite{hwang_direct_2011,liu_visualizing_2011}.
Alternative ways of extracting the topological character of the system
in experiments is, besides the aforementioned Hall effect, are 
Aharonov-Bohm-type interferometry~\cite{duca_aharonov-bohm_2015} in
optical traps and spin-polarized
measurements~\cite{wu_realization_2016}. However, here we focus on
observables that can be easily extended to the transient regime.

\begin{figure}[t]
  \includegraphics[width=\columnwidth]{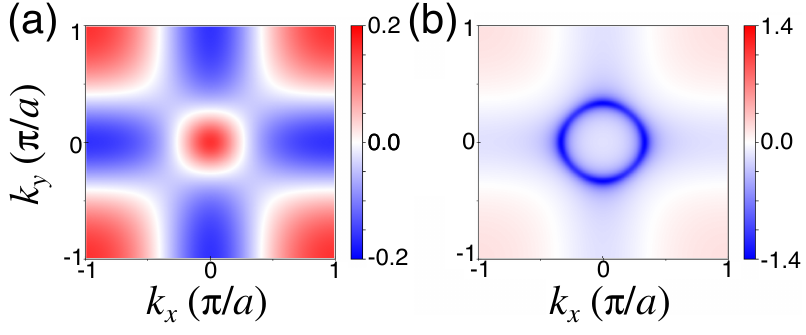}
  \caption{Circular asymmetry $\Delta(\vec{k})$ of the transition
    probability for (a) the BI ($M = M_\mathrm{BI} = -5 |T_0|$) and
    (b) the TI ($M = M_\mathrm{TI} = -3 |T_0|$)
    .\label{fig:asym_matrixelem}
    }
\end{figure}

For the two-band MDM, Fermi's golden rule yields
\begin{align}
  \Gamma^{(\pm)}_1(\omega) \propto \sum_{\vec{k}} \left|\langle \phi_{\vec{k}2} |
  \hat{V}^{(\pm)}(\vec{k}) | \phi_{\vec{k}1} \rangle \right|^2
  \delta(\en_2(\vec{k})-\en_1(\vec{k})-\omega) \ .
\end{align}
Here, the transition operator, derived from the standard Peierls
substitution, reads
\begin{align}
  \hat{V}^{(\pm)}(\vec{k})  = \frac{\partial
  \hat{h}_\mathrm{el}(\vec{k})}{\partial k_x} \pm \iu \frac{\partial
  \hat{h}_\mathrm{el}(\vec{k})}{\partial k_y} \ .
\end{align}
Provided the transition is
permitted by energy conservation, the excitation probability is determined by the matrix elements
$D^{(\pm)}(\vec{k}) = \left|\langle \phi_{\vec{k}2} |
  \hat{V}^{(\pm)}(\vec{k}) | \phi_{\vec{k}1} \rangle \right|^2$. The asymmetry
$\Delta(\vec{k}) =  D^{(+)}(\vec{k})-D^{(-)}(\vec{k})$ is presented for the BI and the TI
in Fig.~\ref{fig:asym_matrixelem} (a) and (b), respectively. We
neglect the el-ph coupling at this point ($\gamma = 0$).

As can be seen in Fig.~\ref{fig:asym_matrixelem}, the main 
difference between BI and TI is the strong negative asymmetry in the
vicinity of the $\Gamma$ point in the latter case (panel (b)),
which is most pronounced 
at the
$\vec{k}$-points where the avoided crossing between the two bands
occurs. This can be understood from the fact that $\Delta(\vec{k})$ is
proportional to the Berry curvature in the two-band
case~\cite{tran_probing_2017}.

\begin{figure}[t]
  \includegraphics[width=\columnwidth]{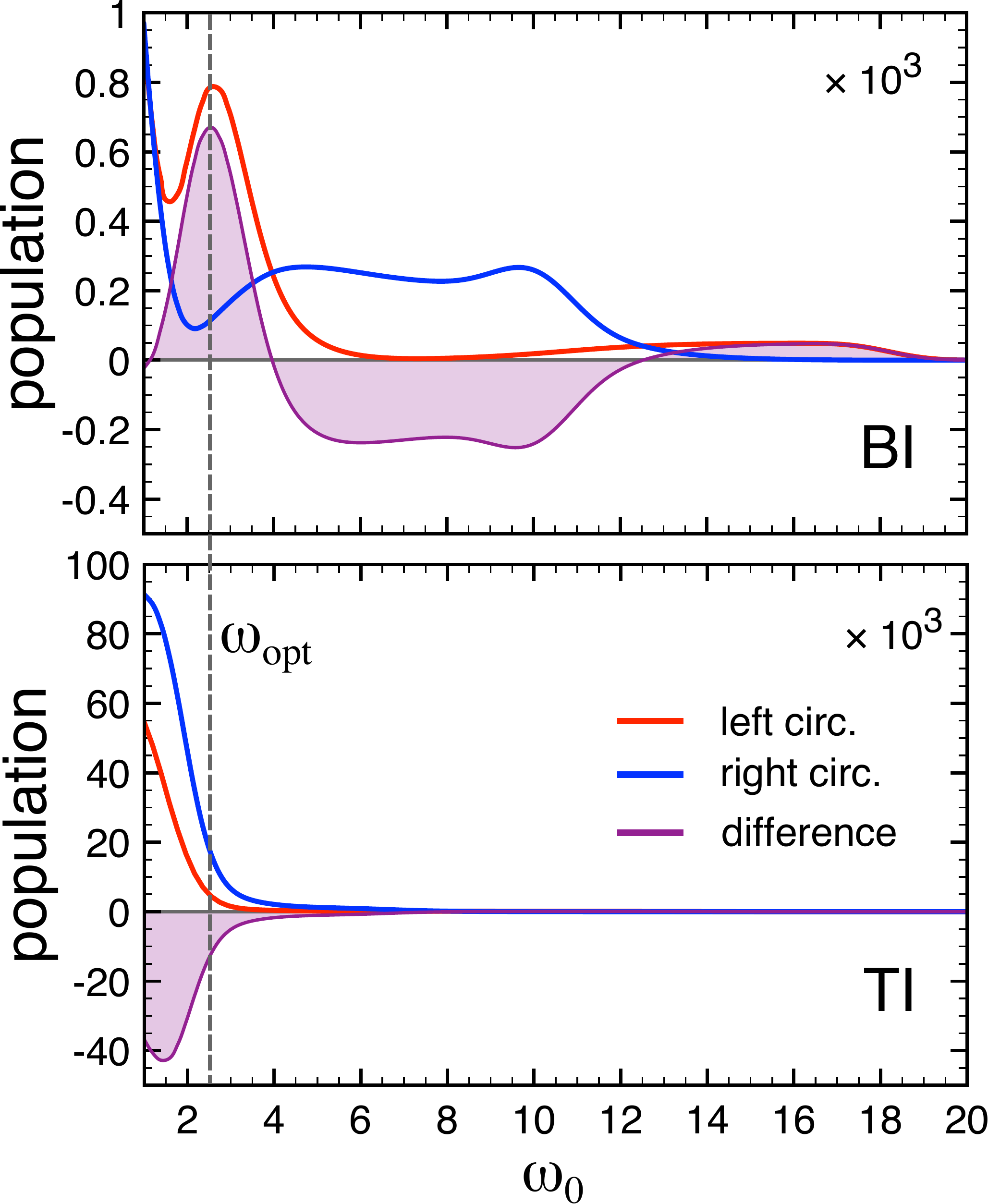}
  \caption{Population of the upper band after applying a pulse with
    central frequency $\omega_0$ to the BI (upper panel) and TI
    (lower panel). The dashed line marks the optimal pulse
    frequency $\omega_\mathrm{opt}$ where the difference between the
    BI and TI response is most pronounced.
    \label{fig:asym_pulse}}
\end{figure}

In view of transient measurements, it would be most useful to
extract the topological character of the system by applying short
(circularly polarized) pulses rather than the continuous-wave (plus
integrating over frequencies) approach presented in 
Ref.~\onlinecite{tran_probing_2017}. Figure~\ref{fig:asym_matrixelem}
suggests to consider transitions close to the $\Gamma$ point, where the
difference in the Berry curvature between the BI and TI is most
pronounced. The thus required spectral resolution has to be balanced
against time resolution determined by the pulse duration. 
We have chosen electric field pulses of the form
\begin{align}
  \label{eq:efield}
  \vec{E}^{(\pm)}(t)=F_0 \sin^2\left(\frac{\pi
  (t-t_0)}{T_{\mathrm{p}}} \right)  \mathrm{Re} \big\{ e^{-\iu
  \omega_0 (t-t_0)}\gvec{\epsilon}^{(\pm)}   \big\} 
\end{align}
for $t_0 < t < t_0 + T_{\mathrm{p}}$. The left or right polarization vector 
(superscript $+$ or $-$, respectively) is given by
$\gvec{\epsilon}^{(\pm)} = \vec{e}_x \pm \iu \vec{e}_y$. In order to
determine the corresponding asymmetry signal, we propagate the
time-dependent Schr\"odinger equation in the presence of the electric
field~\eqref{eq:efield} by the Peierls substitution
$\hat{h}^{(\pm)}_\mathrm{el}(\vec{k};t) \rightarrow
\hat{h}^{(\pm)}_\mathrm{el}(\vec{k}-\vec{A}^{(\pm)} (t);t) $
($\vec{A}^{(\pm)} (t)$ is the vector potential corresponding to the
field) in the weak-field limit ($F_0 = 0.5$). The expected left-right
asymmetry of the depletion rate translates into an asymmetry of the
photoexcitation probability.  After some tests we found that pulses as
short as $T_{\mathrm{p}} = 5.0$ offer a good comprise between
sharpness in frequency space and pulse duration. The excitation
probability for both the BI and the TI case is depicted in
Fig.~\ref{fig:asym_pulse} as a function of the central frequency
$\omega_0$.

In line with Ref.~\onlinecite{tran_probing_2017}, integrating over all
frequencies yields zero for the BI (since $\mathcal{C}_1 = 0$),
while a nonzero value is obtained for the TI. 
As expected from the behavior of the matrix elements (see
Fig.~\ref{fig:asym_matrixelem}), the region in reciprocal space where
the Berry curvature is the strongest 
in the TI case is particularly
suited for mapping out the topological character of the system. This
leads to the optimal frequency $\omega_\mathrm{opt} \simeq 2.5$ where
the BI predominantly absorbs left circularly polarized light
(corresponding to the red region around the origin in
Fig.~\ref{fig:asym_matrixelem}(a)), but where the Berry curvature
leads to a strongly enhanced absorption of right circularly polarized
radiation in the TI case. Choosing 
$\omega_0 = \omega_\mathrm{opt}$ we obtain a field pulse which is ideally suited for tracing the
transient dynamics of the system upon photoexcitation or after a
quench. Note that the absorbance of the BI is reduced compared to the TI 
due to the larger band gap.  

\begin{figure}[ht]
  \includegraphics[width=\columnwidth]{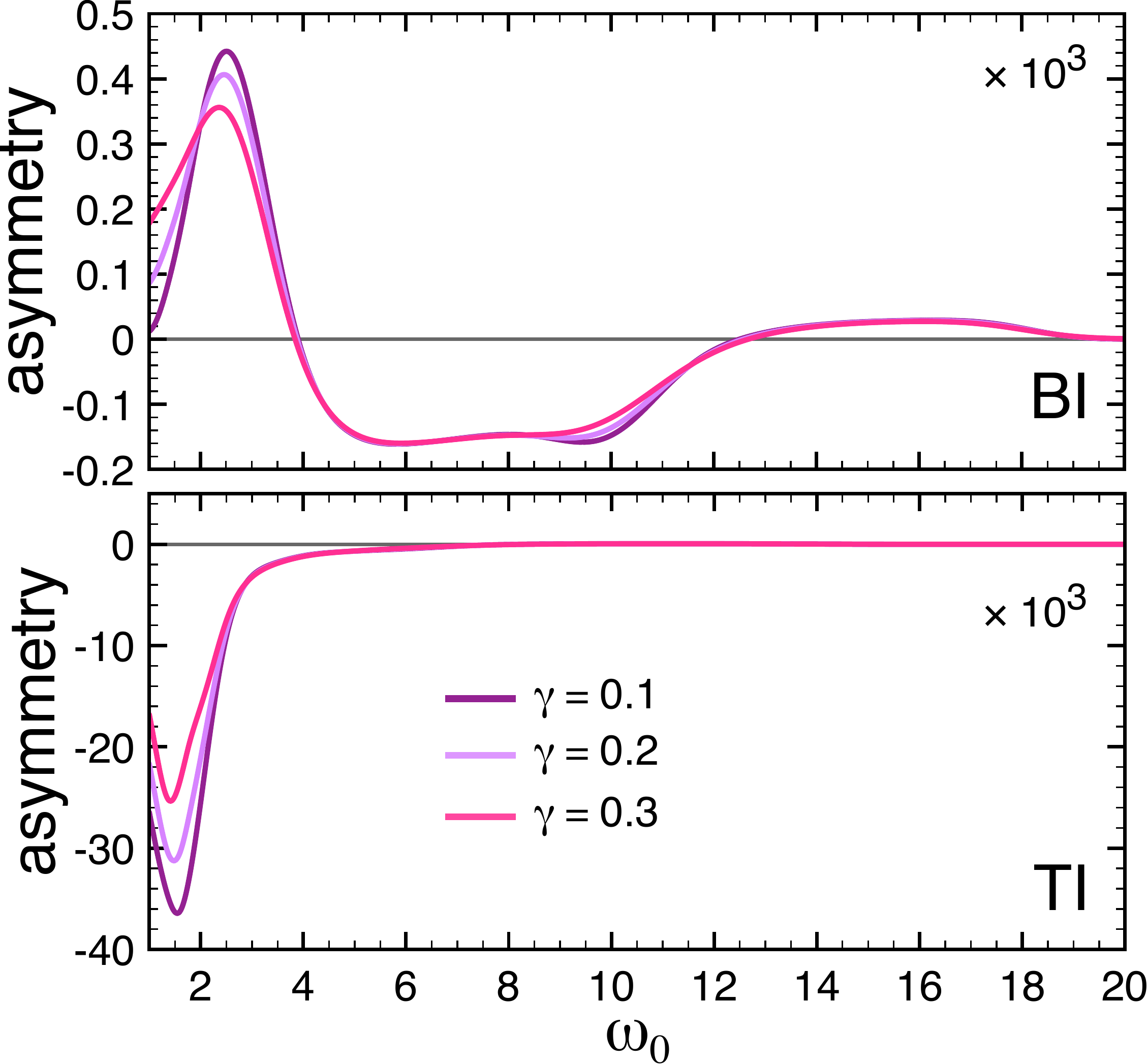}
  \caption{Asymmetry signal analogous to Fig.~\ref{fig:asym_pulse},
    but for different el-ph coupling strengths $\gamma>0$. \label{fig:asym_pulse_phonons}
    \label{fig:asym_pulse_phonons}}
\end{figure}

The next important question to address is if a similar behavior can be
expected in the presence of el-ph coupling. To address this issue, we solved the master
equation~\eqref{eq:eomrho1} including the electromagnetic field and
computed, as for the non-interacting case, the photoexcitation
probability. The result for several moderate coupling strengths $\gamma$ is
presented in Fig.~\ref{fig:asym_pulse_phonons}. Besides an overall
suppression of the absorbance and the less pronounced difference between the 
the excitation probabilities for left/right circularly polarized light,
the qualitative behavior is consistent with the dissipation-less case
(Fig.~\ref{fig:asym_pulse}). The visible difference of the asymmetry
for the relatively strong el-ph couplings demonstrated in
Fig.~\ref{fig:asym_pulse_phonons} 
implies that one can obtain valuable information on the topological character even for
weak to moderate strength of dissipative effects.

\subsection{Time-resolved Hall effect\label{subsec:tdhall}}

\begin{figure}[b]
  \includegraphics[width=\columnwidth]{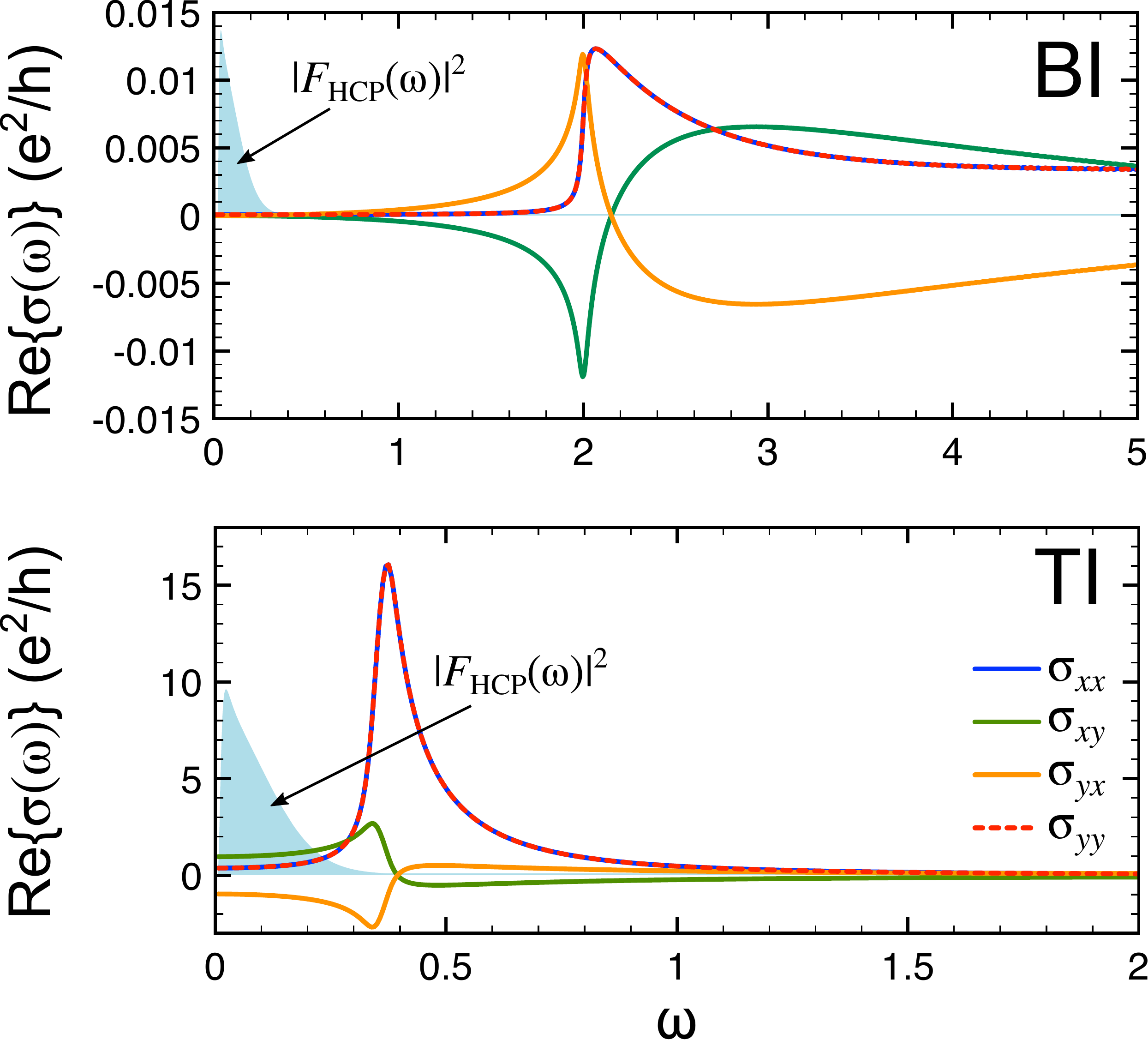}
  \caption{Real part of the optical conductivity for the BI (upper
    panel) and the TI (lower panel). The filled curve illustrates the
    power spectrum of the HCP used for the time-dependent
    calculations (see text). The small but finite value of
    $\sigma_{xx}(\omega = 0)$ is due a small broadening of the Fermi
    surface.  
    \label{fig:reconduct}}
\end{figure}

The emergence of the integer Hall effect $\sigma_{xy} = (e^2/h)
\mathcal{C}_1$ provides direct access to the topological character in
equilibrium. A possible extension of this concept to a time-dependent
scenario is -- similarly as in subsection~\ref{subsec:circasym} -- to
apply suitably shaped pulses with parameters optimized in the
equilibrium case. To this end, we computed the optical conductivity
$\sigma_{\alpha \beta}(\omega)$ for both the BI and the TI phase. The
real part is shown in Fig.~\ref{fig:reconduct}.

The plateau in $\sigma_{xy}(\omega)$ at small frequencies gives us some guidance in choosing the 
spectral features of a suitable probe pulse $\vec{E}(t)$ which allows to
map out the topological character: (i) the pulse needs to be short to
enable us to trace the transient dynamics, (ii) the pulse in frequency
space needs to have a maximal overlap with the region
$\omega \approx 0$, and (iii) $\int\!\dd t\, \vec{E}(t) = 0$ is
required within the dipole approximation. Electromagnetic
field pulses which optimally fulfil the criteria (i)--(iii) are
half-cycle pulses (HCPs)~\cite{moskalenko_charge_2017}. HCPs are
pulses with a dominant, short peak and a weak and long tail
(which hardly influences the dynamics). The dominant peak makes the
field effectively unipolar, as the spectral weight is maximal in the vicinity of
$\omega\approx 0$. 
Here we employ the parameterization $\vec{E}_\mathrm{HCP}(t) = E_0
\gvec{\epsilon} F _\mathrm{HCP}(t-t_0)$ with
\begin{align}
  \label{eq:paramhcp} 
  F_\mathrm{HCP}(t) = x\left(e^{-x^2/2} - \frac{1}{b^2}e^{-x/b } \right) \ , \
 x=\frac{t}{T_\mathrm{p}} \ .
\end{align}
The parameters are the effective pulse duration $T_\mathrm{p}$ and the
shape parameter $b$, which we fix at $b=8$ in
accordance \cite{moskalenko_revivals_2006} with typical pulses generated
in experiments~\cite{you_generation_1993,jones_ionization_1993}.
The analytical expression in Eq.~\eqref{eq:paramhcp} fulfils
$\int^{\infty}_0 \! \dd t \, F_\mathrm{HCP}(t) =0$ exactly.  The
corresponding power spectrum in frequency space is shown in
Fig.~\ref{fig:reconduct}. We have found $T_\mathrm{p}=10$ to be a good
compromise between a short pulse duration and maximal overlap with the
plateau region of the TI, as can be seen in Fig.~\ref{fig:reconduct}.

\begin{figure}[t]
  \includegraphics[width=\columnwidth]{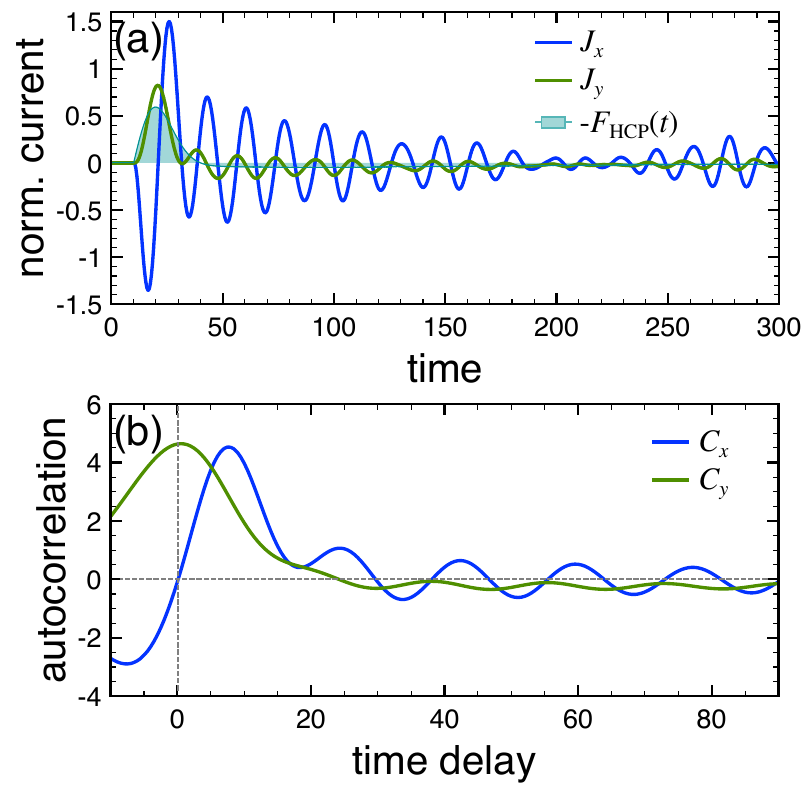}
  \caption{(a) Time dependent current $J_\alpha(t)$ induced by a HCP with
    $T_\mathrm{p}=10$, $t_0 = 10$ and $E_0=10^{-4}$, polarized in the
    $x$-direction. The current has been normalized by the pulse
    strength. (b) Corresponding pulse-current correlation functions
    (Eq.~\eqref{eq:autocorr}) as functions of the delay $\Delta t$.
    \label{fig:tieqcurr}}
\end{figure}

The properties of the HCPs translate -- in combination with the
frequency dependence of the optical conductivity -- into a distinct
behavior of the time-dependent current
\begin{align}
  J_\alpha(t) = \sum_{\vec{k}}
  \mathrm{Tr}\left[\vec{v}^\alpha(\vec{k}-\vec{A}(t)) \gvec{\rho}(\vec{k};t) \right]\ .
\end{align}
Here, $\vec{v}^\alpha(\vec{k})=\partial
\vec{h}_\mathrm{el}(\vec{k})/\partial k_\alpha$ is the velocity matrix.
Within the weak-field regime, linear response theory applies and
relates the current to the driving field 
and the optical conductivity: 
\begin{align}
  \label{eq:linresp1}
  J_\alpha(t) = E_0 \sum_{\beta=x,y} \int^t_{0}\! \dd t^\prime \,
  \sigma_{\alpha \beta}(t-t^\prime) \epsilon_\beta
  F_\mathrm{HCP}(t^\prime -t_0) \ .
\end{align}
Assuming a linearly polarized HCP in the $x$-direction, the linear
response relation~\eqref{eq:linresp1} yields for the current in the 
$y$-direction in frequency space
\begin{align}
  \label{eq:linresp2}
  J_y(\omega) = E_0 \sigma_{y x}(\omega) F_\mathrm{HCP}(\omega)
  \simeq E_0 \sigma_{y x}(\omega=0) F_\mathrm{HCP}(\omega) \ ,
\end{align}
which implies for the time-dependent current
\begin{align}
  \label{eq:linresp3}
  J_y(t) \simeq E_0 \sigma_{y x}(\omega=0) F_\mathrm{HCP}(t)\ .
\end{align}
Thus, the current orthogonal to the field polarization is expected to
closely resemble the driving pulse in the TI case, while the current
will almost vanish for the BI. This behavior is confirmed by the
numerical solution of the Schr\"odinger equation in the presence of the
HCP~\eqref{eq:paramhcp} and the resulting time-dependent current
$J_\alpha(t)$, $\alpha=x,y$ displayed in Fig.~\ref{fig:tieqcurr} for
the TI case. 
The
Hall current $J_y(t)$ has a strong peak for times where the HCP has
its maximum, while the current $J_x(t)$ shows an oscillatory
behavior. As expected, the simulations show that the Hall current is
negligible for the BI phase (not shown).

Since detecting a time-dependent current on the typical time scale of the
pulse (which is in the femto- to picosecond range) is difficult in 
experiments, we propose to analyze the behavior of the
pulse-current correlation function 
\begin{align}
  \label{eq:autocorr} 
  C_{\alpha}(\Delta t) = \int^{\infty}_{0}\!\dd t \, J_\alpha(t)
  F_\mathrm{HCP}(t-\Delta t) \ .
\end{align}
This signal could be detected similarly to the total induced charge,
but weighted with the known driving pulse.
The behavior of the Hall current $J_y(t)$ observed in
Fig.~\ref{fig:tieqcurr} can thus be characterized by a peak at $\Delta
t=0$, while a transient current response originating from a pronounced
variation with respect to the frequency will not possess this
feature. This is confirmed in Fig.~\ref{fig:tieqcurr}.

\section{Tracing the quench dynamics -- unitary time evolution\label{sec:unitdyn}}

Let us now investigate the dynamics of the system after a quench across
the phase boundary and the manifestation of this transition in the observables discussed
in subsection~\ref{sec:obseq}. We focus on the noninteracting case
$\gamma=0$ first.

\begin{figure}[t]
  \includegraphics[width=\columnwidth]{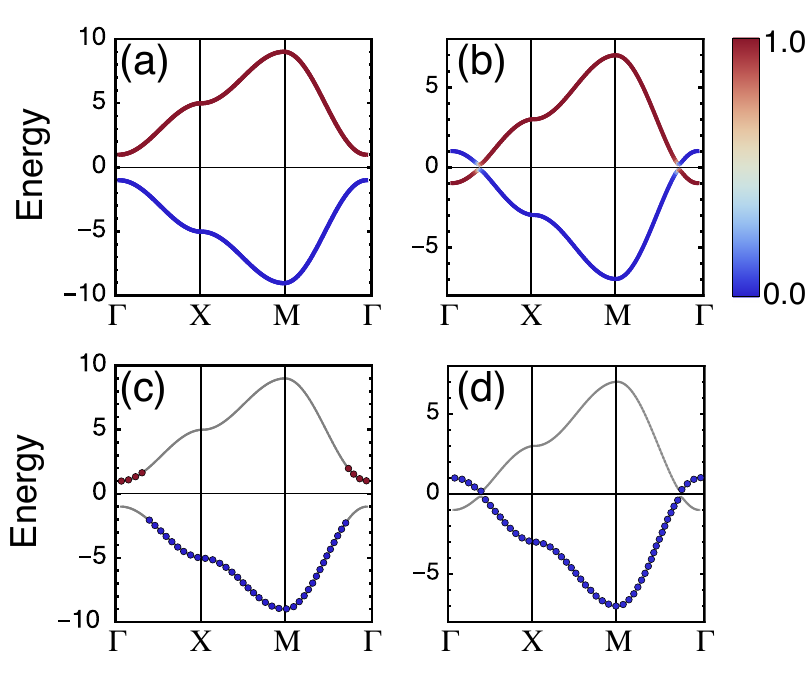}
  \caption{Band structure of (a) the BI and (b) the TI. The
    coloring of the lines indicates the orbital weight of the upper
    band underlying the Hamiltonian~\eqref{eq:Hel}.
 Band structure of (c) the BI and (d) the TI, where 
 the dots indicate that the occupation of the respective
    band after the quench is larger than $0.5$. The energy is measure
    in units of $|T_0|$
    \label{fig:quenchpop}}
\end{figure}

Assuming that the system is initially in equilibrium ($M=M_\mathrm{TI/BI}$)
with the lower band completely filled and the upper band empty,
the gap parameter is ramped up or down in a time 
$T_\mathrm{q}=5$ (see subsection~\ref{subsec:quenchdyn}). This short
ramp time corresponds to an almost ideal quench, i.\,e. the
occupations of the post-quench bands is given by
\begin{align}
  \label{eq:postquenchocc}
  f^{\mathrm{B}}_n(\vec{k}) = \left|\langle
  \phi^{\mathrm{B}}_{\vec{k}n}| \psi_{\vec{k}1}(t=T_\mathrm{q}) \rangle
  \right|^2 \simeq \left|\langle
  \phi^{\mathrm{B}}_{\vec{k}n}| \phi^{\mathrm{A}}_{\vec{k}1} \rangle 
  \right|^2.
\end{align}
The post-quench occupation~\eqref{eq:postquenchocc} is shown along the
standard path in the Brillouin zone of the square lattice in
Fig.~\ref{fig:quenchpop}, marked by points where $
f^{\mathrm{B}}_n(\vec{k}) > 0.5$. In fact, $f^{\mathrm{B}}_n(\vec{k})
$ is close to one for most points. 

\subsection{Time-dependent Hall effect\label{subsec:tdhall_quench}}

\begin{figure}[t]
  \includegraphics[width=\columnwidth]{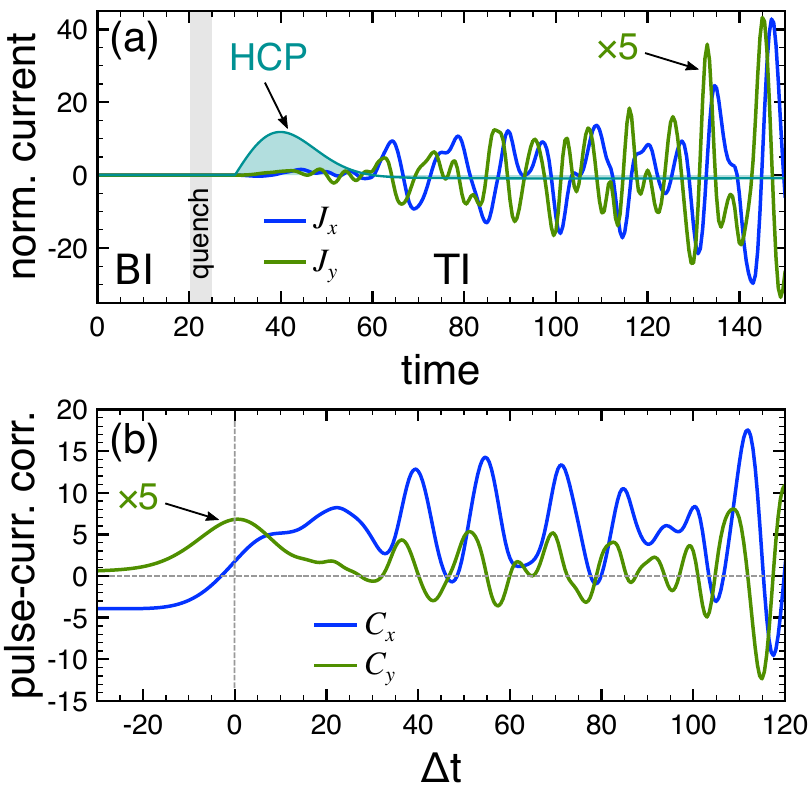}
  \caption{(a) Time-dependent current induced by a HCP (sketched in
    the plot) after the quench from the BI state to the TI
    (illustrated by the shaded background). The current in the $y$
    direction has been multiplied by $5$ for better visibility. The
    normalization of the current is consistent with
    Fig.~\ref{fig:tieqcurr}. (b) Corresponding pulse-current
    correlation functions. 
    \label{fig:quench_bi2ti_curr_corr}}
\end{figure}

Examining the structure of the density matrix after the quench, one
realizes that coherent superpositions of the two bands play a major
role, such that the system is far away from a steady state for which
the Hall conductance~\eqref{eq:hallneq} can be defined. However, to
have a measure of the post-quench state the system is driven to, we
can investigate the response to HCPs as discussed in
subsection~\ref{subsec:tdhall}.  The current induced by a HCP
polarized along the $x$ direction after performing the quench
$M_\mathrm{BI} \rightarrow M_\mathrm{TI}$ is shown in
Fig.~\ref{fig:quench_bi2ti_curr_corr}(a). In comparison to the current
response in equilibrium (Fig.~\ref{fig:tieqcurr}), the coherent
oscillations of the current dominate the hump at times when the
electric field of the pulse is strong. The magnitude of these
oscillations is considerably larger than in the equilibrium case,
showing that the superposition state after the quench is quite
different from the equilibrium TI state. 

At first glance, is seems that the current in the $y$ direction is not
displaying the behavior discussed in subsection~\ref{subsec:tdhall}, 
as the magnitude of the current is quite small even when the
electric field reaches its maximum. Nevertheless, the pulse-current
correlation function $C_{\alpha}(\Delta t)$ defined in 
Eq.~\eqref{eq:autocorr} and presented in
Fig.~\ref{fig:quench_bi2ti_curr_corr}(b), exhibits the distinct
feature of the QHI: $C_y(\Delta t)$ possesses a clear maximum at
$\Delta t =0$, which indicates a plateau behavior of the optical
conductivity $\sigma_{xy}$ at $\omega=0$ and thus the presence of a
nonzero Hall effect. The magnitude of $C_y(\Delta t =0)$ is, however,
significantly (approximately by a factor of four) reduced with respect
to the equilibrium case (Fig.~\ref{fig:tieqcurr}(b)). Hence one would
expect a static Hall conductance of about
$\sigma_{xy}\simeq 0.25 e^2/h$.

We also performed analogous simulations for the $M_\mathrm{TI}
\rightarrow M_\mathrm{BI}$ quench and found the emergence of a very
small, but finite Hall effect after the quench.  

The coherent superposition present in the post-quench state results in
an internal dynamics which might interfere with transient
measurements. However, one can expect decoherence effects to diminish
these coherences, leading to a mixed steady state. In this case,
the static Hall effect yields valuable information on the post-quench
state, as discussed in the next subsection.

\subsection{Steady-state conductance}

\begin{figure}[t]
  \includegraphics[width=\columnwidth]{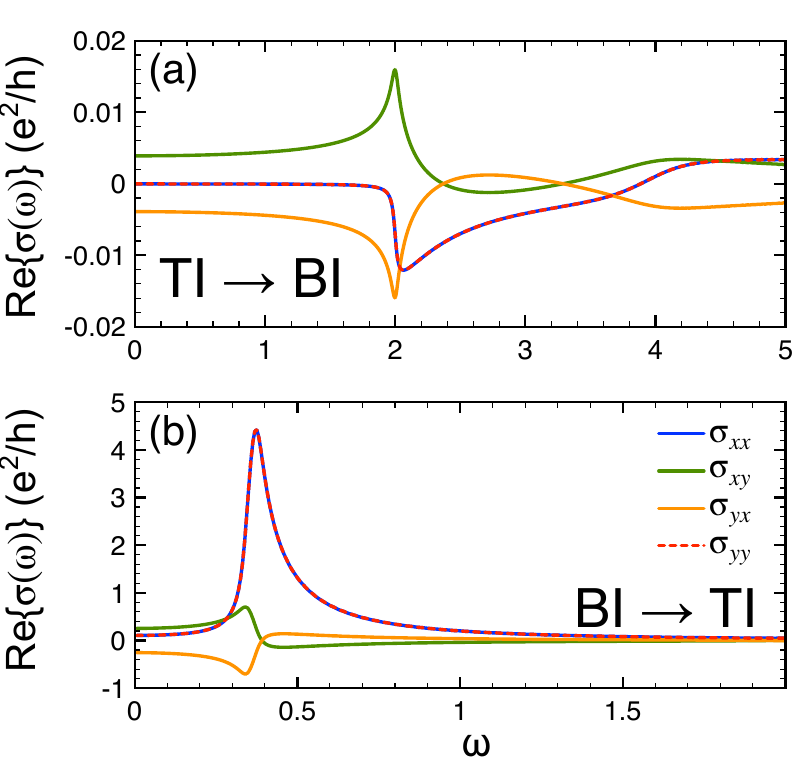}
  \caption{Optical conductivity of the post-quench state
    $\sigma^{\mathrm{B}}_{\alpha \beta}(\omega)$ according to the
    definition~\eqref{eq:neqsigma} for (a) the quench $M_\mathrm{TI}
    \rightarrow M_\mathrm{BI}$ and (b) $M_\mathrm{BI}
    \rightarrow M_\mathrm{TI}$.
    \label{fig:conductquench}}
\end{figure}

As can be inferred from Fig.~\ref{fig:quenchpop}, the occupation after
the quench reflects the pre-quench situation. For instance, the
complete filling of the lower BI band is preserved when switching to the
TI, apart from the avoided crossing points. This illustrates the
conservation of the topological character of the system as the
occupation of bands with the same orbital character (which is
interchanged at the crossing points in the TI band structure, see
Fig.~\ref{fig:quenchpop}) remains constant. Nevertheless, the Hall
effect deviates from the equilibrium behavior, which can be seen by
evaluating the steady-state optical conductivity in accordance with
Ref.~\onlinecite{wang_universal_2016} by
\begin{align}
\label{eq:neqsigma}
  \sigma^{\mathrm{B}}_{\alpha \beta}(\omega) = \frac{1}{\omega}\sum_n
  \int\!\!\dd \vec{k}\, f^\mathrm{B}_n(\vec{k}) \int^\infty_0 \! \dd t
  \, e^{i\omega t} \langle \phi^\mathrm{B}_{\vec{k} n}| \Big[\hat{j}_{\vec{k} \alpha},\hat{j}_{\vec{k} \beta}(t)\Big]|
  \phi^\mathrm{B}_{\vec{k} n} \rangle \ .
\end{align}
Here,
$\hat{j}_{\vec{k} \alpha} = \partial
\hat{h}^\mathrm{B}_\mathrm{el}/\partial k_\alpha$ denotes the
momentum-resolved current operator. The resulting conductivity is
shown in Fig.~\ref{fig:conductquench}. Note that computing the optical
conductivity analogously to Eq.~\eqref{eq:hallneq} assumes that all
off-diagonal elements of the density matrix, which capture the
coherent oscillations of the system after the excitation, are zero.

As can be inferred from Fig.~\ref{fig:conductquench}, the Hall
conductance significantly deviates from the equilibrium value. In the
$M_\mathrm{TI} \rightarrow M_\mathrm{BI}$ quench, the system acquires a finite
Hall conductance
$\sigma^{\mathrm{B}}_{xy}(\omega=0) \simeq 0.0038 e^2/h$, while the
quench $M_\mathrm{TI} \rightarrow M_\mathrm{BI}$ leads to
$\sigma^{\mathrm{B}}_{xy}(\omega=0) \simeq 0.25 e^2/h$. These values
are consistent with the time-dependent response discussed in
subsection~\ref{subsec:tdhall_quench}. Without the unit factor $e^2/h$, the latter can be
regarded as a nonequilibrium generalization of the Chern
number~\cite{wang_universal_2016}. Note that the concrete numbers
depend on both the pre- and the post-quench gap parameter $M$.

\subsection{Transient circular asymmetry\label{subsec:tdasym_unitquench}}

Let us now proceed to transient properties. As discussed in
subsection~\ref{subsec:circasym}, the circular asymmetry is a very
promising candidate for tracing the dynamics in real time. To find a
suitable analogue to the equilibrium scenario, we performed test
calculations of the population dynamics driven by circularly polarized
pulses (with the same parameters as in
subsection~\ref{subsec:circasym} and with central frequency
$\omega_0=\omega_\mathrm{opt}$) after the system has been
quenched. For the case $M_\mathrm{BI} \rightarrow M_\mathrm{TI}$ one
finds that the system is preferably excited by right circular pulses,
while a left circular pulse results in a weaker depletion of the
post-quench lower band. In contrast, circularly polarized pulses
applied to the BI after the quench from the TI result in a depletion
of the \emph{upper} band instead (irrespective of the
polarization). Furthermore, one observes an oscillatory dependence of the
photoexcitation probability after a quench, which originates --
analogously to the current discussed in
subsection~\ref{subsec:tdhall_quench} -- from the coherent
superposition of the states belonging to the upper and lower band,
respectively. 

\begin{figure}[t]
  \includegraphics[width=\columnwidth]{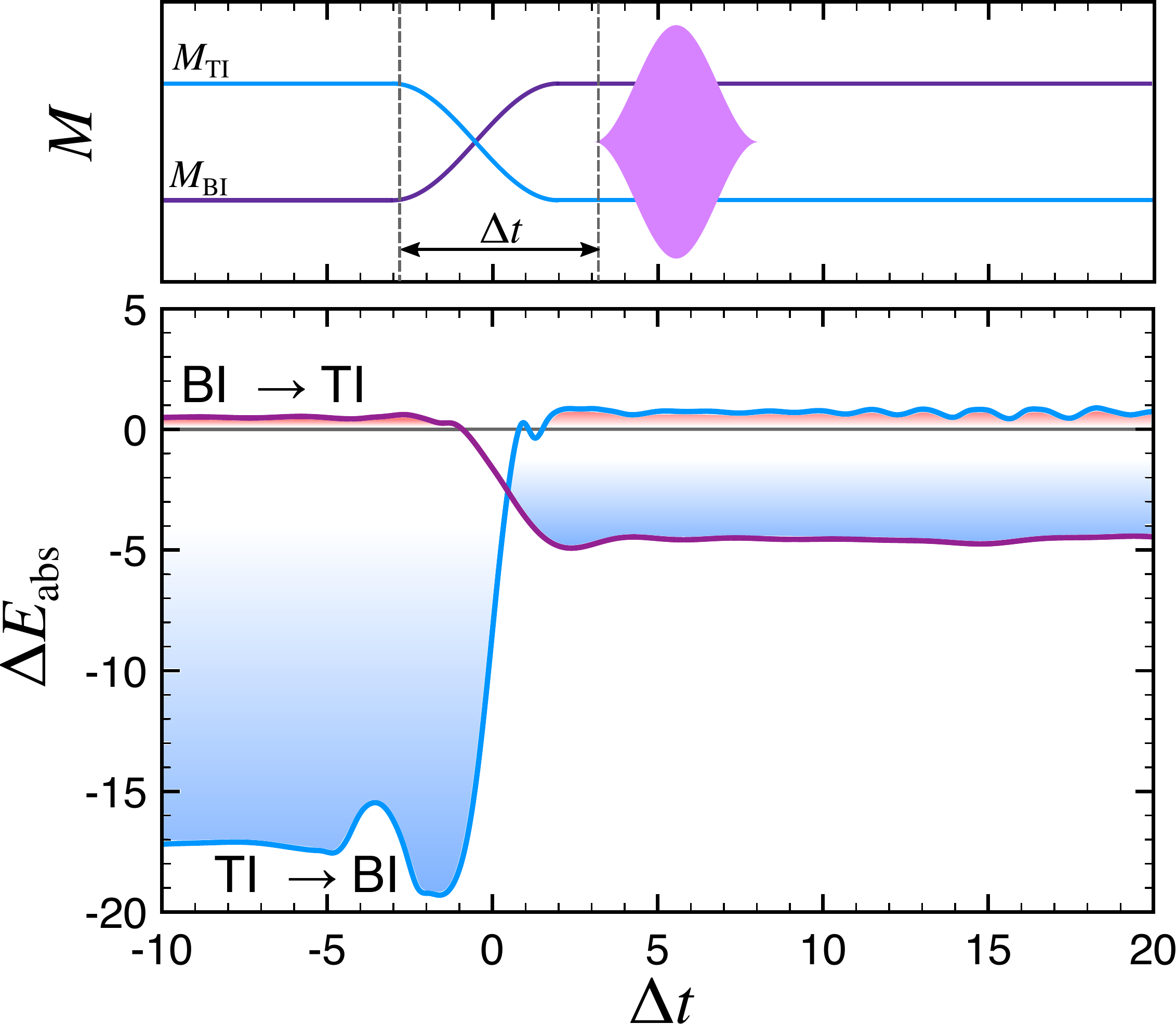}
  \caption{The circular asymmetry of the absorbed energy $\Delta
    E_\mathrm{abs}$ 
    as a function of the delay $\Delta t$ (lower panel) between the quench and the probe pulse (illustrated in the
    upper panel). The color gradient shading of the curves emphasizes 
    the transition from positive (red) to negative (blue) asymmetry.
    \label{fig:tdasym}}
\end{figure}

For these reasons, we propose to utilize the absorbed \emph{energy}
$E_\mathrm{abs}$ of the left or right circular probe pulses as a
footprint of the topological character in nonequilibrium instead. Importantly,
the energy absorption can be measured in experiments directly by
placing photon detectors behind the sample.
It should be noted that $E_\mathrm{abs}$ can also be negative when the
system is excited, which corresponds to stimulated emission rather
than absorption. This difference to the equilibrium case needs to
be taken into account. We thus define a transient generalization of
the circular asymmetry by
\begin{align}
  \label{eq:tdasym}
  \Delta E_\mathrm{abs}= \left| E^{(+)}_\mathrm{abs}\right|
  - \left| E^{(-)}_\mathrm{abs}  \right|
\end{align}
where $E^{(\pm)}_\mathrm{abs}$ is the energy of the probe pulse which
is absorbed (or emitted) by the system. The time-resolved circular
asymmetry signal~\eqref{eq:tdasym} for both quench scenarios is
presented in Fig.~\ref{fig:tdasym} as a function of the delay
$\Delta t = t_0 - t_\mathrm{q}$ between the starting time of the pulse
($t_0$) and the time when the system is quenched ($t_\mathrm{q}$). The
absorbed energy is computed by taking the energy difference
$E^{(\pm)}_\mathrm{abs} = E^{(\pm)}_\mathrm{q+pr}
-E^{(\pm)}_\mathrm{q}$, where $E^{(\pm)}_\mathrm{q}$ denotes
the total energy of the quenched system in absence of the probe pulse,
while $E^{(\pm)}_\mathrm{q+pr}$ is the total energy of the quenched
and probed target. The energy is measured at a sufficiently large
reference time after the quench and the pulse.

Figure~\ref{fig:tdasym} demonstrates that the quench dynamics can be
traced in the time domain by the circular asymmetry. For the system with $M=M_\mathrm{BI}$
before switching, the asymmetry is positive (see
Fig.~\ref{fig:asym_pulse}). Hence, for $\Delta t<0$ one observes
$\Delta E_\mathrm{abs}(T_\mathrm{ref}) >0$ in Fig.~\ref{fig:tdasym} for the
$M_\mathrm{BI}\rightarrow M_\mathrm{TI}$ case (purple curve). For
$\Delta t >0$, the asymmetry assumes negative values (with larger
modulus, as well) which -- in accordance to Fig.~\ref{fig:asym_pulse}
-- indicates the transition to the TI. 
The opposite behavior can be
observed when switching $M_\mathrm{TI}\rightarrow M_\mathrm{BI}$ (blue
curve). 
Further features are weakly pronounced coherent oscillations of the
asymmetry after the quench to the TI which originate from the
off-diagonal elements of the time-dependent density matrix. The
corresponding time scale is determined by the energy difference
between the states whose occupation is changed by the quench.

\begin{figure*}
  \includegraphics[width=0.9\textwidth]{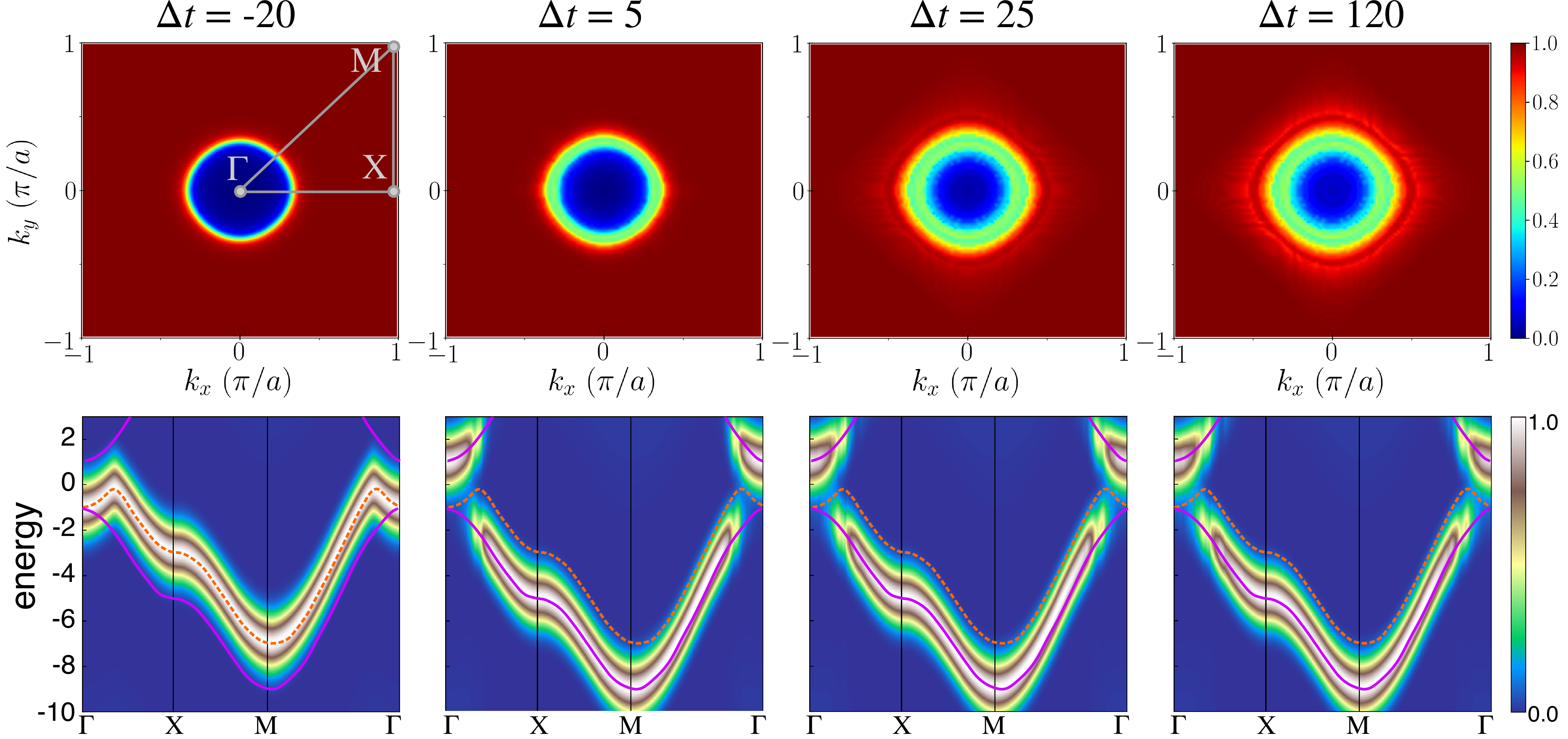}
  \caption{Upper row: time-dependent occupation with respect to the
    post-quench Hamiltonian (defined by $M=M_\mathrm{BI}$) for
    different time delays $\Delta t$ relative to the quench time (at $\Delta t =0$). 
    Lower row: corresponding tr-ARPES spectra according
    to Eq.~\eqref{eq:photocurr1}. The full purple lines indicate the
    post-quench BI band structure, while the dashed orange lines is
    the lower band of the TI. The el-ph coupling is set to $\gamma =
    0.5$.
    \label{fig:pop_pe_ti2bi}}
\end{figure*}

As our time-dependent simulations demonstrate, the time-resolved
quench-probe asymmetry signal based on the absorbed energy provides a
robust tool to trace the transient dynamics of the system after a
quench. It primarily maps out the circular asymmetry of the underlying
bands and is less sensitive to the nonequilibrium occupation. These
features clearly distinguish the time-resolved asymmetry from the
nonequilibrium Hall effect discussed above and render it a powerful
complementary tool.

\section{Tracing the quench dynamics -- dissipative time
  evolution\label{sec:elphdyn}}

After having analyzed the unitary dynamics of the system after a
quench, we now investigate how the picture changes if el-ph
interactions, as discussed in subsection~\ref{subsec:elph}, are
present. Generally, the effect of coupling to the phonon modes is
expected to give rise to dissipative dynamics, lowering the energy
after the quench excitation. Revisiting Fig.~\ref{fig:quenchpop} one
can expect a qualitatively different behavior for the quench
$M_\mathrm{TI} \rightarrow M_\mathrm{BI}$ as compared to the case
$M_\mathrm{BI} \rightarrow M_\mathrm{TI}$. If the system is quenched
from the TI to the BI, the occupation in the upper band (see 
Fig.~\ref{fig:quenchpop}(a)) is located around the energy minimum at
the $\Gamma$ point. Hence, no energy can be extracted from the system
after the quench. The effect of the el-ph coupling is in this case primarily 
the dephasing of the coherences induced by the quench.

\subsection{Transient dynamics probed by time-resolved photoemission
\label{subsection:trarpes}}

We performed numerical simulations of the quench dynamics by solving
the master EOM~\eqref{eq:eomrho1} as described in
subsection~\ref{subsec:eomelph}. The parameters are -- apart from the
el-ph interaction -- the same as in
section~\ref{sec:unitdyn}. To understand the time evolution of
the band structure and the nonequilibrium occupation, the most
convenient quantity to look at is the time-dependent occupation with
respect to the post-quench basis
$f^{\mathrm{B}}_n(\vec{k};t) = \rho_{nn}(\vec{k};t)$ and -- as
complementary information -- the transient photoelectron
spectrum. Time-resolved ARPES
(tr-ARPES) has recently become a standard tool for tracing the time
evolution in correlated
systems \cite{schmitt_transient_2008,graf_nodal_2011,
  bovensiepen_elementary_2012,smallwood_tracking_2012}, in parallel
with a rapid development of state-of-the-art theoretical descriptions
within the NEGF
framework~\cite{sentef_examining_2013,kemper_mapping_2013,murakami_damping_2016}. Modeling
the photoemission process by a pump pulse (frequency $\omega$, pulse
envelope $F_\mathrm{pr}(t)$) yields the photocurrent~\cite{kemper_mapping_2013}
\begin{align}
  \label{eq:photocurr1}
  I(\vec{k};\omega) \propto \mathrm{Im}\int^\infty_{-\infty}\! \dd t
  \! \int^t_{-\infty}\! \dd t^\prime &  \, \left(F_\mathrm{pr}(t)
  \right)^*F_\mathrm{pr}(t^\prime) e^{-\iu \omega (t-t^\prime)}
                      \nonumber \\ & \times 
  \mathrm{Tr}\left[\vec{G}^<(\vec{k};t,t^\prime) \right] \ .
\end{align}
Here, $\vec{G}^<(\vec{k};t,t^\prime)$ denotes the lesser Green's function,
which we express within the generalized Kadanoff-Baym ansatz as
$\vec{G}^<(\vec{k};t,t^\prime) = \iu \vec{U}(\vec{k};t)\gvec{\rho}(\vec{k};t^\prime)$ with the
time-evolution operator corresponding to the single-particle
Hamiltonian $\vec{h}_\mathrm{el}(\vec{k};t)$. Varying the delay
$\Delta t$
between the excitation (the quench at $t_\mathrm{q}$ in our case) and
the time when the probe pulse is applied, Eq.~\eqref{eq:photocurr1}
provides a $\vec{k}$- and energy-resolved pump-probe spectrum. 

\begin{figure*}
  \includegraphics[width=0.9\textwidth]{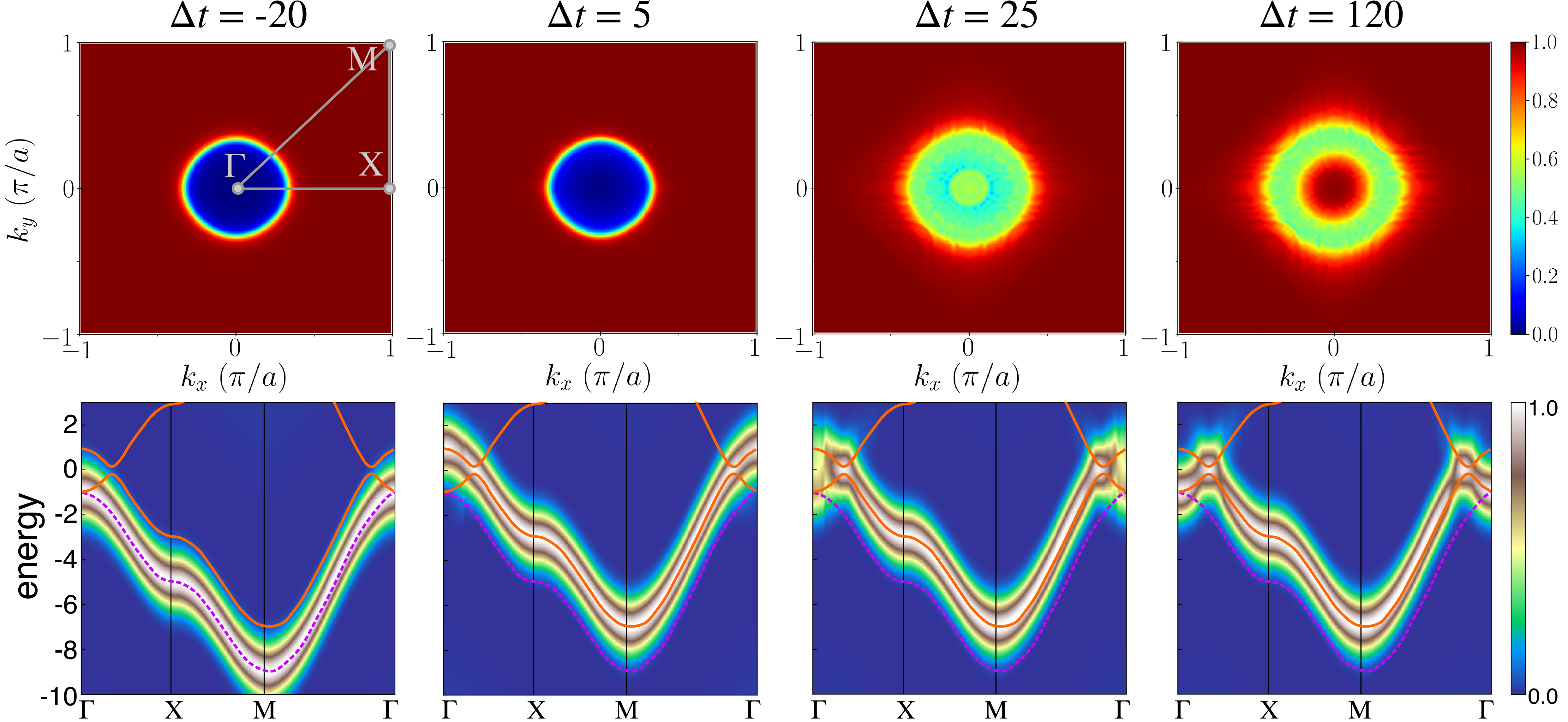}
  \caption{Upper row: time-dependent occupation with respect to the
    post-quench Hamiltonian (defined by $M=M_\mathrm{TI}$) for
    different time delays $\Delta t$ as in Fig.~\ref{fig:pop_pe_ti2bi}. 
    Lower row: tr-ARPES spectra according to
    Eq.~\eqref{eq:photocurr1}. The full purple lines indicate the
    post-quench TI band structure, while the dashed orange line represents 
    the lower band of the BI. The el-ph coupling is set to $\gamma =
    0.5$.
    \label{fig:pop_pe_bi2ti}}
\end{figure*}

In Fig.~\ref{fig:pop_pe_ti2bi} we present the time-dependent
occupation $f^{\mathrm{B}}_n(\vec{k};t)$ (upper row) 
along with the
corresponding tr-ARPES spectra (lower row) as a function time delay
$\Delta t$ for the quench $M_\mathrm{TI} \rightarrow M_\mathrm{BI}$. In the calculations of 
the tr-ARPES signals, we used the same parameters for the
pulse as in section~\ref{sec:unitdyn}. The short pulse length
$T_\mathrm{p}=5$ gives rise to the broadening of the spectra in
Fig.~\ref{fig:pop_pe_ti2bi}.
At $\Delta t = -20$, the system is still in equilibrium and the ARPES
signal follows the pre-quench band structure (orange dashed line). The
post-quench basis is the BI -- therefore, the occupation with respect
to the BI bands exhibits a hole around the $\Gamma$ point up to the region
where the avoided crossing occurs. This demonstrates a different aspect of the topological
state: the occupation of the first Brillouin zone with respect to the dominant orbital character of the
lower band is not singly-connected. Following the time evolution right
after the quench ($\Delta t=5$) we see a similar picture as in
Fig.~\ref{fig:quenchpop}: the ARPES spectrum now follows the
post-quench band structure (solid purple lines), while the upper band
is populated around the $\Gamma$ point, whereas the lower band is
empty in this region. The population assumes values between zero and
one around the crossing region, such that a slight lowering of the
total energy by el-ph relaxation is possible. This effect can be
observed for later times ($\Delta t=25$). The steady state ($\Delta
t=120$) however shows only a slight blurring of the
occupation compared to directly after the quench.
 
The nonequilibrium dynamics is considerably more pronounced in the quench
scenario $M_\mathrm{BI} \rightarrow M_\mathrm{TI}$, analyzed again in
terms of the time-dependent population of the post-quench lower band in 
the tr-ARPES spectra in Fig.~\ref{fig:pop_pe_bi2ti}.

\begin{figure}
  \includegraphics[width=\columnwidth]{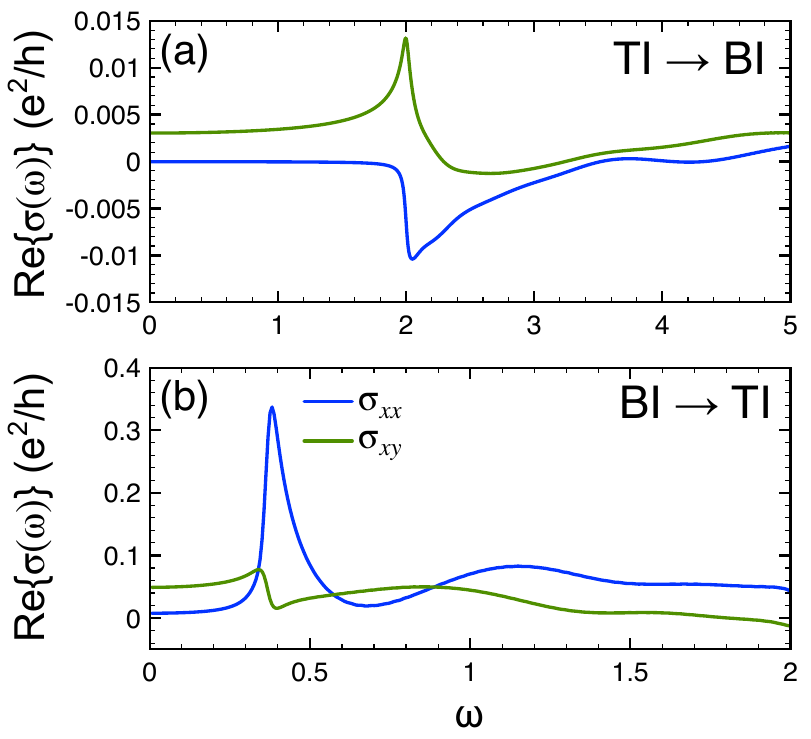}
  \caption{Real part of the nonequilibrium optical conductivity of
    the post-quench steady-state reached in the presence of
    el-ph interactions, for (a) the transition $M_\mathrm{TI}
    \rightarrow M_\mathrm{BI}$, and (b) the switching $M_\mathrm{BI} \rightarrow M_\mathrm{TI}$.
    \label{fig:conductivity_re_quench_ph}}
\end{figure}

Right after the quench ($\Delta t =5$), the ARPES spectrum closely
resembles the BI band structure, apart from a shift to larger
energies. However, as the occupation of the lower band shows, the
post-quench TI band is empty between the $\Gamma$ point and the
crossing region, while the upper TI band is populated in this region
in the Brillouin zone. It is clear from Fig.~\ref{fig:pop_pe_bi2ti} that 
this nonequilibrium occupation does not correspond to an energy
minimum as filling the occupation hole in the lower band and a
relaxation towards the energy minimum at the crossing points in the
upper band result in a lowering of the total electronic energy. These 
dissipation processes are efficiently mediated by the el-ph
interaction as the time-dependent occupation and the ARPES spectra for
later times ($\Delta t=25$) demonstrate. The steady state reached at
$\Delta t = 120$ has a peculiar configuration: the occupation hole
around the $\Gamma$ point has been filled, while the population of the
upper band has relaxed down to the crossing point. Interestingly, from
the ARPES spectrum alone one could suspect that the system has fully
relaxed to a TI. However, the nonequilibrium occupation which involves
both bands gives rise to a steady-state optical conductivity
(Fig.~\ref{fig:conductivity_re_quench_ph}) which deviates considerably
from the equilibrium behavior (Fig.~\ref{fig:reconduct}) in its
strongly suppressed conductance and Hall conductance. The direct current Hall conductance is reduced to
$\sigma_{xy}\simeq 0.003 e^2/h$ for the quench to the BI, while
we find $\sigma_{xy}\simeq 0.05 e^2/h$ in the
$M_\mathrm{BI} \rightarrow M_\mathrm{TI}$ scenario. Note that the
value of the Hall conductance in the TI final state is considerably
smaller than in the quench scenario without el-ph
interaction (Fig.~\ref{fig:conductquench}). This can be explained by
the distinct occupation in the post-quench steady-state in the presence of
the el-ph coupling. The integer Hall conductance $\sigma_{xy}=e^2/h$
originates from interband transitions in the crossing region of the upper
and lower band. Exactly those transitions are strongly suppressed in
the steady-state of Fig.~\ref{fig:pop_pe_bi2ti}, as the occupation in
the lower band in the crossing region is depleted, while the
available states in the upper band are mostly occupied.

\subsection{Transient circular asymmetry}

\begin{figure}[t]
  \includegraphics[width=\columnwidth]{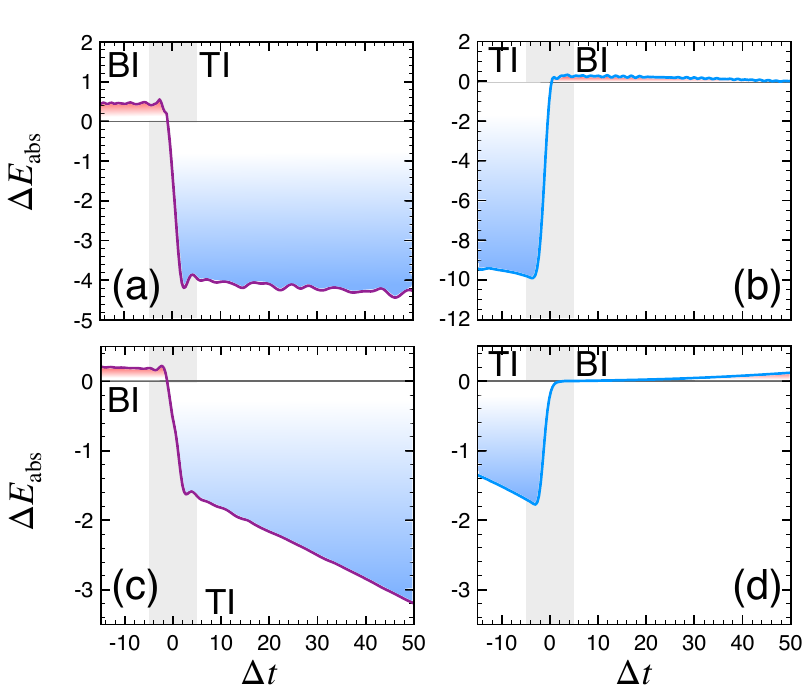}
  \caption{Asymmetry of the absorbed energy $\Delta E_\mathrm{abs}$
    as a function of the delay $\Delta t$ between the quench
    and the probe pulse for (a)--(b) weak el-ph coupling $\gamma=0.1$,
    and (c)--(d) moderate el-ph coupling $\gamma=0.3$. The color
    gradient filling of the curves is analogous to
    Fig.~\ref{fig:tdasym}. The gray shaded area
    indicates the time when the system is switched from BI to TI
    (left panels) or TI to BI (right panels), respectively.
    \label{fig:quench_tdasym_ph}}
\end{figure}

Let us now investigate if the transition from the topologically trivial BI
to the QHI or vice-versa can be traced in the time domain by
circularly polarized pulses in an analogous fashion as for the
dissipation-less case discussed in subsection~\ref{subsec:tdasym_unitquench}. We
employ the same scheme: a left or right circularly polarized pulse
photoexcites the system before, during or after the quench. The
energy absorbed by the pulse and, in particular, the left-right asymmetry
should then reveal the character of the steady state or transient
state. We have applied this recipe, using the
definition~\eqref{eq:tdasym}, and present the respective
energy absorption asymmetry in Fig.~\ref{fig:quench_tdasym_ph} for two
representative cases of the el-ph interaction: (i) weak coupling
$\gamma=0.1$ (Fig.~\ref{fig:quench_tdasym_ph}(a)--(b)) and moderate
coupling strength $\gamma=0.3$
(Fig.~\ref{fig:quench_tdasym_ph}(c)--(d)). 
We represent
the absorption asymmetry as a function of the quench-pulse delay $\Delta t$ as in
Fig.~\ref{fig:tdasym}. 

For $\gamma=0.1$, Fig.~\ref{fig:quench_tdasym_ph}(a) shows a
transition $M_\mathrm{BI} \rightarrow M_\mathrm{TI}$ similarly to the
unitary case. The asymmetry $\Delta E_\mathrm{abs}$ changes from positive values
(with small magnitude) to negative, indicating the switch from BI to
TI. On the other hand, the quench
$M_\mathrm{TI} \rightarrow M_\mathrm{BI}$
(Fig.~\ref{fig:quench_tdasym_ph}(b)) is accompanied by a switch of
$\Delta E_\mathrm{abs}$ from negative to positive 
values (with strongly reduced magnitude). Generally,
the scheme for tracing the quench dynamics works very well for the
dissipative case with weak el-ph interaction.

Turning to stronger el-ph couplings, the picture slightly changes. For the
quench $M_\mathrm{BI} \rightarrow M_\mathrm{TI}$
(Fig.~\ref{fig:quench_tdasym_ph}(c)), one can observe the
asymmetry $\Delta E_\mathrm{abs}$ switching from small positive values to the
negative region across the quench. The almost linear decrease of
$\Delta E_\mathrm{abs}$ towards larger $\Delta t$ is attributed to the el-ph
induced relaxation after the pulse (see
Fig.~\ref{fig:pop_pe_bi2ti}). One can readily check that the
photoexcitation probability due to the specific pulse is larger in the
post-quench relaxed steady state than right after the quench, because
more states are occupied in the energy window defined by the pulse
frequency. Therefore, the absorption of the left-circular probe pulse
becomes more efficients as $\Delta t$ increases, until the steady
state is reached (at $\Delta t \sim 200$). Generally, the dependence
of $\Delta E_\mathrm{abs}$ on $\Delta t$ is much more pronounced in
the TI phase, since the influence of the el-ph coupling (as discussed
in subsection~\ref{subsection:trarpes}) is stronger. 

\section{Conclusions}

We have studied the quench dynamics of the MDM as a generic model for
two-dimensional topological insulators with a special emphasis on how
the nonequilibrium and transient properties are reflected in
experimentally accessible quantities. We have focused on two promising
observables which reveal the topological character of the system: the
steady-state and time-dependent Hall effect and the asymmetry of
photoexcitation with respect to left or right circularly polarized
pulses. Based on a realistic model for two-dimensional topological
insulators we have defined suitable probe-pulse shapes by considering
the equilibrium model, both for the dissipation-less case and
including electron-phonon interactions. We then applied these
optimized pulses to trace the nonequilibrium dynamics after a quench.
Both the time-dependent Hall effect and the circular dichroism of the
absorbed energy provide valuable information on the system. While the
Hall current can deliver important insights into the steady state, it
turns out to be less suitable for the analysis of transient
states. This is due to the coherent superpositions in the system,
which give rise to an intrinsic transient dynamics. The circular
asymmetry of the absorbed energy, on the other hand, is much less
sensitive to these effects, since it is based on the occupation of the
bands only. The latter approach is thus particularly well suited for
the study of the switching process between different phases.

In the presence of electron-phonon coupling, the quench dynamics can
significantly differ from the dissipation-less case, provided the
energy of the system is reduced by scattering from phonons. We
analyzed these effects in terms of the time-resolved ARPES spectra. We
investigated the steady-state properties by the nonequilibrium Hall
conductance and, finally, analyzed the quench dynamics in terms of the
circular asymmetry. As an important point for the potential
experimental realization, we have demonstrated the robustness of our
prosed transient measurement in the presence of weak to moderate
electron-phonon coupling. The transient circular dichroism has thus
been shown as a very promising tool to obtain insights in the up to
now little explored field of nontrivial topological phases in
nonequilibrium.

\appendix

\section{Numerical solution of the master equation \label{app:mastereq}}

For a stable numerical solution of a master equation of the type of
Eq.~\eqref{eq:eomrho1} we adapt a method for propagating NEGFs in
time~\cite{stan_time_2009}. The interval $[0,t_\mathrm{max}]$ is
discretized into an equidistant grid $t_n=n \Delta t$. In order to
perform the step $\gvec{\rho}(t_n) \rightarrow \gvec{\rho}(t_{n+1})$,
we separate the unitary time evolution from the scattering term by
the ansatz
\begin{align}
  \label{eq:rhosep1}
  \gvec{\rho}(t_n+\tau) &= \vec{U}(t_n+\tau,t_n)
  \widetilde{\gvec{\rho}}_n(\tau) \vec{U}^\dagger(t_n+\tau,t_n)
                          \nonumber \\
  &\equiv \vec{U}_n(\tau) \widetilde{\gvec{\rho}}_n(\tau)
  \vec{U}^\dagger_n(\tau) \ , \tau \in [0,\Delta t] \ .
\end{align}
Here, $\vec{U}(t_n+\tau,t_n)$ denotes the time-evolution operator,
which we approximate by $\vec{U}_n(\tau) \approx \exp[-\iu\, \tau
\vec{h}_\mathrm{el}(t_n + \Delta t/2)]$. Inserting
Eq.~\eqref{eq:rhosep1} into the EOM~\eqref{eq:eomrho1} then yields 
\begin{align}
  \label{eq:rhoprop1}
  \gvec{\rho}(t_{n+1}) &= \vec{U}_n(\Delta t) \gvec{\rho}(t_{n})
                         \vec{U}^\dagger_n(\Delta t) \nonumber \\ &\quad+ \int^{\Delta
                         t}_0 \! \dd \tau \,
                  \vec{U}^\dagger_n(\tau-\Delta t) \vec{I}(t_n+\tau)
                                                                    \vec{U}_n(\tau-\Delta
                                                                    t)
                                                                    \ .
\end{align} 
Apart from the approximation to the time-evolution operator
$\vec{U}_n(\tau)$, Eq.~\eqref{eq:rhoprop1} is still exact. A simple
and numerically stable propagation scheme is obtained by approximating
$\vec{I}(t_n+\tau)\approx \vec{I}(t_n + \Delta t/2)\equiv
\vec{I}_{n+1/2}$. Using the Baker-Hausdorff formula, the time
step~\eqref{eq:rhoprop1} can be expressed as 
\begin{align}
  \label{eq:rhoprop2}
  \gvec{\rho}(t_{n+1}) &=  \vec{U}_n(\Delta t) \gvec{\rho}(t_{n})
                         \vec{U}^\dagger_n(\Delta t) \nonumber \\
                       &\quad+\iu \Delta t
                         \sum^{p-1}_{k=0}  \frac{(-\iu \Delta t)^k}{k!}
                         \vec{C}^{(k)}_n + \mathcal{O}\left((\Delta t)^{p+1}\right)
\end{align}
with $\vec{C}^{(0)}_n=\vec{I}_{n+1/2}$ and
$\vec{C}^{(k+1)}_n=\big[\vec{h}_\mathrm{el}(t_n + \Delta t/2),
\vec{C}^{(k)}_n \big]$. In practice, we truncate
Eq.~\eqref{eq:rhoprop2} after the fourth order ($p=4$). The half-step
scattering term $\vec{I}_{n+1/2}$ is obtained by fourth-order
polynomial interpolation using $\vec{I}_{n+1-k}$, $k=0,\dots,3$. This
requires knowing the scattering term at the next time step,
$\vec{I}_{n+1}$, which is a function of the yet unknown density matrix
$\gvec{\rho}(t_{n+1})$. Therefore, we employ a predictor-corrector
scheme where $\vec{I}_{n+1/2}$ is first estimated by third-order
polynomial extrapolation, allowing to compute $\gvec{\rho}(t_{n+1})
$ and thus obtain $\vec{I}_{n+1}$. The latter two steps are then
iterated at each time step until $\gvec{\rho}(t_{n+1})$ is converged.

\begin{acknowledgments}
  The calculations have been performed on the Beo04 cluster at the
  University of Fribourg. This work has been supported by the Swiss
  National Science Foundation through NCCR MARVEL and ERC Consolidator
  Grant No.~724103. We thank Markus Schmitt and Stefan Kehrein for
  fruitful discussions.
\end{acknowledgments}

\vfill

%



\end{document}